\documentclass[journal]{IEEEtran}

\IEEEoverridecommandlockouts                              

\title{
Adaptive Evolutionary Framework for Safe, Efficient, and Cooperative Autonomous Vehicle Interactions
}

\newif\ifblind

\ifblind
    \author{Anonymous Authors}
\else
\author{
    Yijun Lu$^{1}$,
    Zhen Tian$^{2}$,
    and Zhihao Lin$^{2}$%
    \thanks{$^{1}$Yijun Lu is with Waseda University, Tokyo, Japan
    (e-mail: yijun@ruri.waseda.jp).}%
    \thanks{$^{2}$Zhen Tian and Zhihao Lin are with the University of Glasgow, Glasgow, United Kingdom
    (e-mail: tian.zhen@glasgow.ac.uk; 2800400L@student.gla.ac.uk).}%
}

\usepackage{xcolor}

\usepackage{hyperref}
\usepackage{cite}
\usepackage{algorithm}  
\usepackage{algpseudocode}
\usepackage{multirow}
\usepackage{amsmath}

\usepackage{amssymb}
\usepackage{epsfig} 

\usepackage{amsthm}

\usepackage{graphicx}
\usepackage{caption}
\usepackage{subcaption}
\usepackage{enumitem} 

\usepackage[flushleft]{threeparttable}

\usepackage{tabularx} 

\usepackage{titlesec}

\begin{document}

\maketitle

\begin{abstract}
Modern transportation systems face significant challenges in ensuring road safety, given serious injuries caused by road accidents. The rapid growth of autonomous vehicles (AVs) has prompted new traffic designs that aim to optimize interactions among AVs. However, effective interactions between AVs remains challenging due to the absence of centralized control. Besides, there is a need for balancing multiple factors, including passenger demands and overall traffic efficiency. Traditional rule-based, optimization-based, and game-theoretic approaches each have limitations in addressing these challenges. Rule-based methods struggle with adaptability and generalization in complex scenarios, while optimization-based methods often require high computational resources. Game-theoretic approaches, such as Stackelberg and Nash games, suffer from limited adaptability and potential inefficiencies in cooperative settings. This paper proposes an Evolutionary Game Theory (EGT)-based framework for AV interactions that overcomes these limitations by utilizing a decentralized and adaptive strategy evolution mechanism. A causal evaluation module (CEGT) is introduced to optimize the evolutionary rate, balancing mutation and evolution by learning from historical interactions. Simulation results demonstrate the proposed CEGT outperforms EGT and popular benchmark games in terms of lower collision rates, improved safety distances, higher speeds, and overall better performance compared to Nash and Stackelberg games across diverse scenarios and parameter settings.
\end{abstract}


\section{INTRODUCTION}

\IEEEPARstart{A}{s} modern transportation systems present a unique set of challenges and risks, significantly impacting road safety. In Great Britain alone, there were over 1,700 fatalities and more than 29,000 serious injuries reported in 2022~\cite{UKGov2023}. Globally, road traffic crashes claim approximately 1.19 million lives each year~\cite{UKGovRoadCasualties2022}. These records reflects the critical need for improved safety measures in on-road driving. With the fast growth in autonomous vehicles (AVs), projected to be over 50 million by 2024~\cite{ignatious2022overview}, modern traffic designs are evolving to better accommodate AVs. These updates address critical challenges in ensuring safe control and interaction between AVs~\cite{deluka2018introduction, ferrarotti2024autonomous}. Furthermore, AVs have the potential to significantly reduce safety concerns caused by human errors such as distraction, and delayed reactions~\cite{BADUE2021113816}. Additionally, AVs compute optimal decision-making solutions faster than human drivers, enhancing reaction to emergency situations~\cite{inproceedings}. In recent years, the AVs experience fast development and promote the driving safety and traffic efficiency~\cite{parekh2022review,9802527,lin2024conflicts}.  

AVs include three core stages for stable and safe driving, including the perception, decision-to-planning, and control. It is challenging to achieve an effective driving between AVs without a centralized guiding system. Without a centralized guiding system, the interaction between AVs can be complex due to the demand of maximizing their own profits. Additionally, AVs must be capable of balancing multiple demands of passengers, such as safety, efficiency~\cite{ma2023distributed,bichiou2018developing}. Some decision-making frameworks optimize traffic flow and improve safety~\cite{hang2020integrated,9745461,benloucif2019cooperative,10107652,wu2023integrated}. Besides, the AVs is expected to be unselfish, which means they need to consider befitting the general traffic. Therefore, interactive driving between AVs requires balanced decision-making capabilities to avoid collisions, maintain high driving efficiency, and positive effects to publication. Some approaches are used for AVs' interaction, such as rule-based methods, optimization-based methods, and game-based methods.

\begin{figure}[t]
    \centering
    \includegraphics[width=1\linewidth]{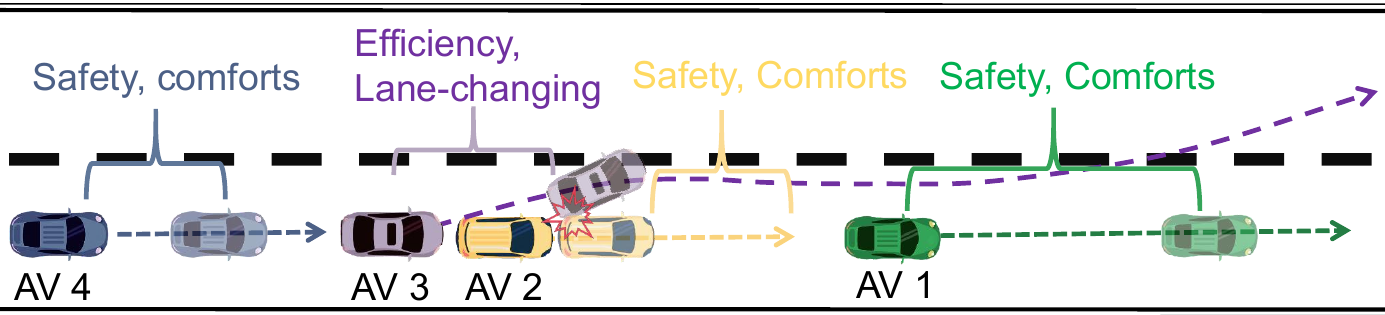}

    \caption{Unsafe interactions between AVs due to their varied demands.}
    \label{fig1}
\end{figure}
Fig.~1 illustrates the complexity of interactive driving between AVs due to differing demands and the challenges of balancing these demands. In this scenario, there are four AVs, each with its own specific demands, such as driving safety, efficiency, comfort, and lane-changing intentions. However, conflicts can arise between different AVs. For instance, prioritizing safety for AV 3 might result in slower driving speeds, while AV 1 focuses on efficiency and attempts to change lanes, potentially leading to a collision in the future. Therefore, effective decision-making frameworks are essential to balance these demands and ensure safe traffic interactions.

Rule-based methods are among the earliest approaches used for controlling autonomous vehicles. These methods rely on predefined sets of rules to govern vehicle behavior with high transparency~\cite{xiao2021rule}. However, the rule-based methods encounter the difficulty of generalization and adaptation. For instance, a rule-based planner is defined in~\cite{bouchard2022rule}, which commands AVs to stop when meet interaction and spot-lines simultaneously. However, this rule is not suitable for the similar scenarios, such as roundabouts with spot-lines. Besides, creating the rules demands substantial engineering expertise and is challenging, which restricts the use of rule-based methods in more complex scenarios. For instance, a Finite State Machine (FSM) is utilized to select driving maneuvers based on comprehensive driving rules for various possible situations~\cite{9729796}. However, these detailed driving rules necessitate accounting for every possible situation with corresponding actions, as well as verifying the state against these rules at each step. Meeting these requirements involves covering all driving scenarios, achieving high precision in state observation, and ensuring high computational efficiency of the devices involved.

Optimization-based methods involve solving mathematical optimization problems to determine the best course of action for a vehicle. Common optimization-based methods include nonlinear programming (NLP), quadratic programming (QP), and model predictive control (MPC). For example, MPC is widely used to optimize vehicle trajectories over a future horizon by minimizing a cost function~\cite{diehl2009efficient}. However, MPC requires real-time optimization, posing cost challenges for practical applications. Hardware-in-the-loop simulations are needed to confirm that MPC is capable of meeting real-time requirements before it can be deployed in real-world settings~\cite{vajedi2015ecological}. QP and NLP are commonly used to determine optimal trajectory parameters \cite{10048483, 10285563}. They also provide flexibility in handling diverse constraints and objectives. However, these methods often involve high computational complexity.

Game-based approaches are well-suited for decision-making in interactive driving scenarios. Two widely used game-based models are Nash and Stackelberg games. Stackelberg games are hierarchical models often used in scenarios involving a leader-follower dynamic. This approach has applications in autonomous vehicle platooning, where a leader vehicle dictates strategies that follower vehicles adopt. For example, interactions between lane-changing vehicles and their adjacent counterparts have been modeled using a Stackelberg game framework~\cite{ding2019multivehicle}. Such models enable adjustment of driving behaviors by varying emphasis on safety, efficiency, and comfort~\cite{hang2020integrated,hang2021decision}. However, Stackelberg games heavily rely on accurate predictions of surrounding vehicles driving styles, leading to significant errors when predictions are inaccurate. Additionally, Stackelberg games lack adaptability and flexibility in real driving, as real-world driving styles can deviate considerably from predefined styles.

Nash games, on the other hand, represent a non-cooperative game theory approach, where each agent selects its optimal strategy given the strategies of others. This approach has been applied to autonomous driving for optimizing lane-changing and merging scenarios~\cite{hang2020human,lin2019pay}. However, Nash equilibrium tends to prioritize self-interest, which can lead to situations where collective efficiency is not guaranteed. For instance, in competitive merging, vehicles may become overly cautious, causing congestion. Therefore, an effective game-based approach is needed, one that avoids constructing a leader-follower relationship and eliminates the need to explicitly identify leader driving styles. Such an approach should maintain the benefits for all vehicles, ensuring strong general performance.

\begin{figure*}[t]
    \centering
    \includegraphics[width=1\linewidth]{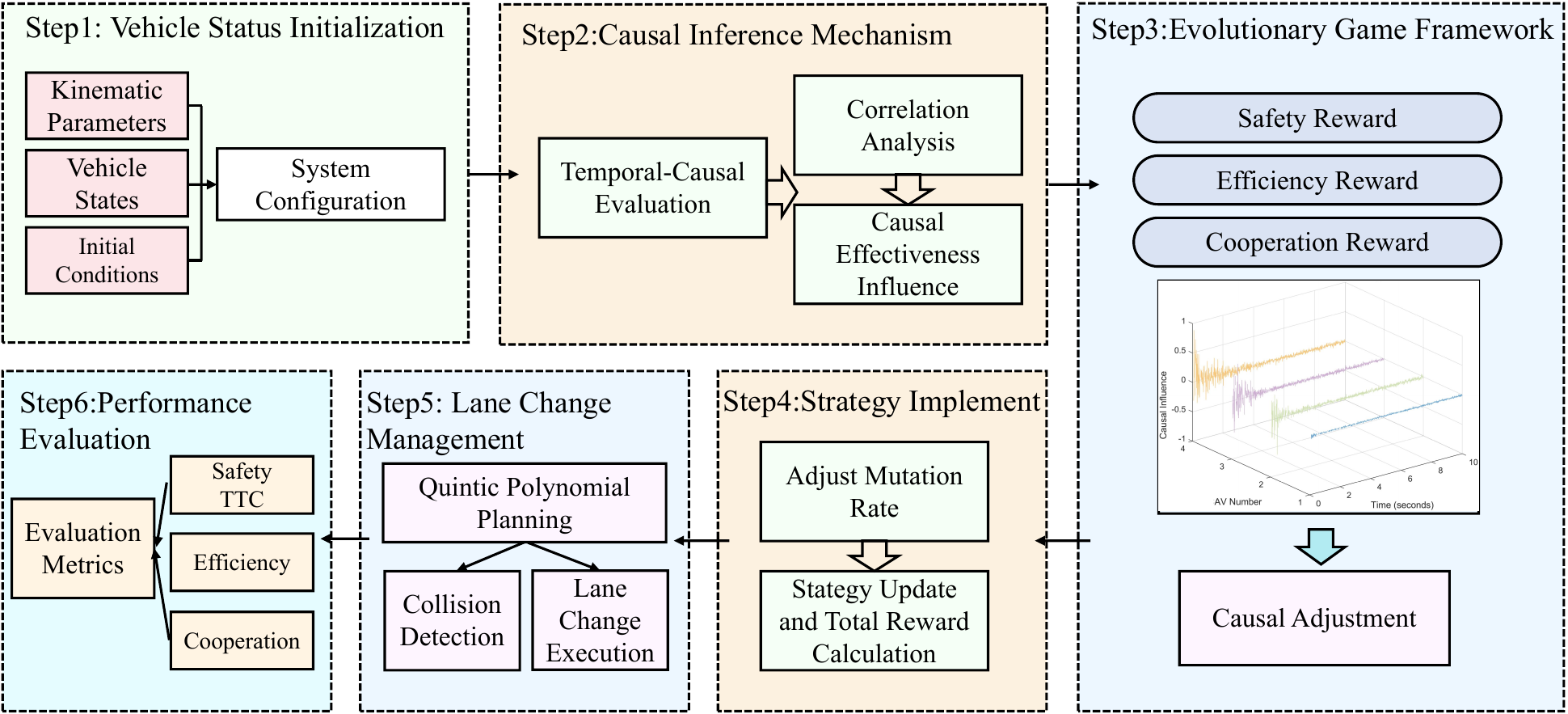}
    \caption{The causal evaluation-based evolutionary game theory for safe, efficienct, and cooperative interactive driving.}
    \label{fig2_interactive}
\end{figure*}
Recently, evolutionary game theory (EGT) has emerged as a promising alternative, utilizing a decentralized, adaptive framework where vehicles evolve their strategies based on past experiences~\cite{traulsen2023future,ahmad2023applications}. Unlike the previously mentioned approaches, EGT does not rely on pre-established rules or hierarchical leadership, making it particularly suitable for dynamic and complex environments. EGT consists of two main modules: evolution and mutation. Historical experience is used to inform decisions at each step. When satisfactory decisions for the current situation exist, the optimal decision from history is adopted as a mutation. When no satisfactory historical decision is available, alternative decisions are explored through evolution. An evolutionary rate is crucial for balancing mutation and evolution. A major advantage of EGT in the context of interaction driving is its ability to dynamically adjust strategies in response to specific tasks. Compared to the Nash game, EGT enables vehicles to continually evolve strategies based on a combination of historical rewards and evolutionary rates. By guiding evolutions toward maintaining higher general benefits, overall performance can be ensured. In contrast to the Stackelberg game, EGT does not rely on responses to a leader's states, allowing all vehicles to operate on equal footing.

Although EGT shows significant advantages over Nash and Stackelberg games, a critical challenge lies in defining an appropriate evolutionary rate to balance mutation and evolution. A fixed evolutionary rate may lead to incorrect evolutions in stable situations, while random changes in the evolutionary rate can result in unintended evolutions. To address this, a causal evaluation module is introduced in this paper to dynamically adjust the mutation rate. This causal evaluation module assesses the relationship between historical performance and strategies, and when strategies result in poor performance, it encourages evolution.

The main contributions of this paper are summarized as follows:
\begin{itemize}
    \item An evolutionary game-theoretic framework is proposed, allowing AVs to adapt their strategies dynamically. This EGT framework improves overall performance by eliminating the leader-follower relationship found in Nash and Stackelberg games.
    \item A causal evaluation module is integrated with EGT (CEGT), enhancing each vehicle's ability to learn from historical interactions. CEGT reduces total collisions and achieves higher total rewards compared to standard EGT across various driving scenarios and parameter settings.
    \item The proposed CEGT effectively avoids collisions, maintains safe relative distances, sustains high average speeds, and achieves high total rewards. CEGT outperforms both Nash and Stackelberg games in terms of reduced total collisions and higher total rewards across multiple scenarios and parameter settings.
\end{itemize}
The rest of this paper is organized as follows: Section~\ref{sec2} describes the problem statement and system structure; Section~\ref{sec3} describes risk-aware and MPC-based trajectory optimization; Section~\ref{sec4} describes the simulation results and analysis; Section~\ref{sec5} draws the conclusions.

\section{System Structure}
\label{sec2}
In order to achieve safe decision making of interactive driving between AVs, a CEGT framework is proposed, as shown in Fig.~2.
The proposed framework leverages a causal-based Evolutionary Game Theory (EGT) approach to optimize AVs interactions. The process begins with vehicle status initialization, where the kinematic parameters, vehicle states, and initial conditions are configured to simulate realistic driving environments. Next, in causal inference mechanism, a temporal-causal evaluation is conducted using correlation analysis to quantify causal effectiveness and influence. The causal influence provides insights into how each AV's actions impact system-level outcomes. In evolutionary game framework, rewards are computed based on safety, efficiency, cooperation, and causal adjustment. These rewards are used to guide strategy evolution, balancing cooperative and individual objectives. Causal adjustments are made iteratively to refine strategies and ensure system-wide optimization. Afterwards, strategy implementation stage adjusts the mutation rate dynamically and updates strategies based on the calculated total reward, ensuring adaptability and optimal performance across varying scenarios. During the lane change management, quintic polynomial planning are used for smooth and safe transitions between lanes. Collision detections are used to ensure the safety during lane-changing. Finally, key metrics such as safety, efficiency, and cooperation are assessed to validate the effectiveness of the framework. Safety is partially reflected by the variations in time-to-collision (TTC) for each AV in this paper.

\section{Methodology} 
\label{sec3}
Our methodology is based on combining causal inference mechanisms and evolutionary game theory for multi-agent decision-making in a dynamic driving scenario.

\subsection{Vehicle Kinematic}
The vehicle kinematics are modeled by considering the longitudinal and lateral dynamics of each vehicle. The position of each vehicle \(i\) at time step \(t\), denoted by \(x_i(t)\), is updated based on its velocity \(v_i(t)\) and the time increment \(\Delta t\):
\begin{equation}
x_i(t+1) = x_i(t) + v_i(t) \cdot \Delta t.
\end{equation}
The velocity of each vehicle is updated based on its current strategy and acceleration \(a_i(t)\):
\begin{equation}
v_i(t+1) = v_i(t) + a_i(t) \cdot \Delta t,
\end{equation}
where the acceleration \(a_i(t)\) is determined by the selected strategy, with safety-focused strategies generally having lower accelerations, while efficiency-focused strategies have higher accelerations.

\subsection{Causal Inference Mechanism}
\begin{figure}[t]
    \centering
    \includegraphics[width=1\linewidth]{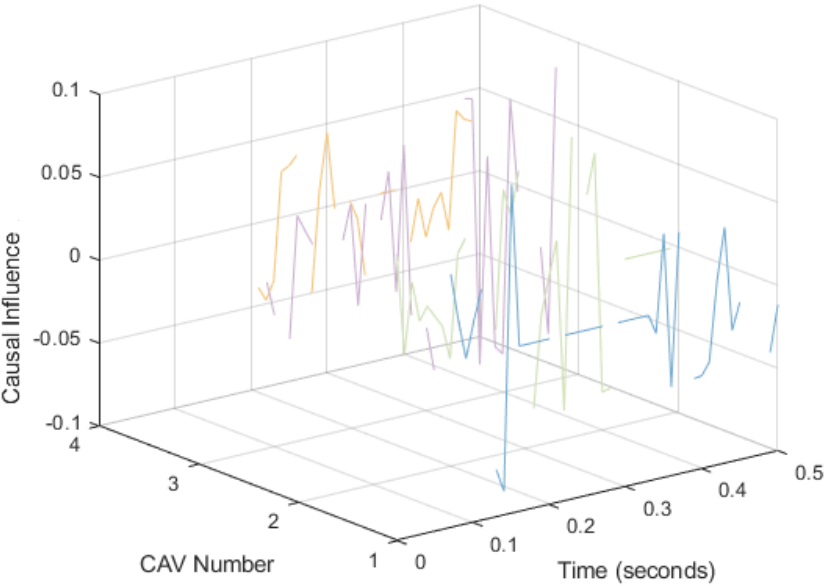}
    \caption{An example of causal influence. }
\end{figure}

To effectively model the complex temporal and spatial dependencies in multi-vehicle systems, we introduce a sophisticated causal inference mechanism that dynamically evaluates and adjusts vehicle behaviors. This mechanism captures both the immediate interactions between vehicles and the historical influence of past decisions on current behaviors.

The causal effectiveness at time step $t$, denoted as $E_{\text{causal}}(t)$, is formulated through a temporal-spatial evaluation framework:
\begin{equation}
E_{\text{causal}}(t) = \frac{\overline{r_s} \cdot (\text{cm} + \overline{C}(t))}{\overline{R_{\text{total}}}}
\end{equation}
Where $\overline{r_s}$ represents the average step reward across all vehicles at time $t$, calculated as $\overline{r_s} = \frac{1}{N}\sum_{i=1}^N r_{s,i}(t)$, $\text{cm}$ is a causal mask parameter that introduces controlled stochasticity to prevent over-exploitation of historical patterns, $\overline{C}(t)$ denotes the mean historical causal influence at time $t$, computed as $\overline{C}(t) = \frac{1}{N}\sum_{i=1}^N C_i(t)$, and $\overline{R_{\text{total}}}$ represents the average cumulative rewards over previous time steps, defined as 
\begin{equation}
\overline{R_{\text{total}}} = \frac{1}{t-1}\sum_{\tau=1}^{t-1} R_{\text{total}}(\tau)
\end{equation}
The causal influence for the multi-vehicle system is represented in matrix form through a correlation-based analysis:
\begin{equation}
\mathbf{C}(t) = w_c \begin{bmatrix}
\text{corr}(\mathbf{P}_1(t), \mathbf{r}_1(t)) \\
\text{corr}(\mathbf{P}_2(t), \mathbf{r}_2(t)) \\
\vdots \\
\text{corr}(\mathbf{P}_N(t), \mathbf{r}_N(t))
\end{bmatrix}
\end{equation}
where $\mathbf{P}_i(t)$ represents the normalized position history vector of vehicle $i$ up to time $t$, while $\mathbf{r}_i(t)$ is the corresponding reward vector. The correlation function $\text{corr}(\cdot)$ measures the statistical dependency between historical positions and current rewards, defined as
\begin{equation}
\begin{aligned}
\text{corr}(\mathbf{P}_i(t), \mathbf{r}_i(t)) &= \frac{\text{cov}(\mathbf{P}_i(t), \mathbf{r}_i(t))}{\sigma_{\mathbf{P}_i(t)} \cdot \sigma_{\mathbf{r}_i(t)}} \\
&= \frac{\sum_{k=1}^{t-1}(\Delta p_i(k))(\Delta r_{s,i}(k))}{\sqrt{\sum_{k=1}^{t-1}(\Delta p_i(k))^2} \sqrt{\sum_{k=1}^{t-1}(\Delta r_{s,i}(k))^2}}
\end{aligned}
\end{equation}
Where where $\sigma_{\mathbf{P}_i(t)} = \sqrt{\sum_{k=1}^{t-1}(\Delta p_i(k))^2}$ is the standard deviation of normalized positions, $\sigma_{\mathbf{r}_i(t)} = \sqrt{\sum_{k=1}^{t-1}(\Delta r_{s,i}(k))^2}$ is the standard deviation of rewards. The correlation measure provides a value in the range [-1, 1], indicating the strength and direction of the relationship between historical positions and current rewards. the position and reward deviations, $\Delta p_i(k)$ and $\Delta r_{s,i}(k)$ are defined as:
\begin{equation}
\Delta p_i(k) = \frac{p_i(k)}{t} - \frac{1}{t-1}\sum_{j=1}^{t-1}\frac{p_i(j)}{t}
\end{equation}
\begin{equation}
\Delta r_{s,i}(k) = r_{s,i}(t) - r_{s,i}(t)
\end{equation}
 $\text{cov}(\mathbf{P}_i(t), \mathbf{r}_i(t))$ represents the covariance between position history and rewards
The position and reward histories are structured as time-series vectors:
\begin{equation}
\mathbf{P}_i(t) = \begin{bmatrix}
p_i(1)/t \\
p_i(2)/t \\
\vdots \\
p_i(t-1)/t
\end{bmatrix}, \quad
\mathbf{r}_i(t) = \begin{bmatrix}
r_{s,i}(t) \\
r_{s,i}(t) \\
\vdots \\
r_{s,i}(t)
\end{bmatrix}
\end{equation}
Where $p_i(k)$ denotes the position of vehicle $i$ at time step $k$, normalized by the current time step $t$ to ensure comparable scales across different time horizons. The reward vector $\mathbf{r}_i(t)$ is replicated to match the dimension of the position history, enabling correlation analysis. The weight coefficient $w_c$ scales the magnitude of causal influence to balance it with other components in the decision-making process. This causal inference framework enables vehicles to learn from their historical experiences while maintaining adaptability to current situations through the stochastic component introduced by the causal mask parameter. Fig.~2 illustrates an example of causal influences for four CAVs. For each CAV, the causal influence fluctuates between -0.1 and 0.1 within a 0.5-second period. This suggests that during this time period, the causal influences of the four CAVs are constantly seeking a balance between imitation and mutation within the evolutionary game.

\subsection{Evolutionary Game Theory}
The evolutionary game framework is formulated as a multi-agent decision process where vehicles adapt their strategies through a sophisticated balance of imitation and mutation. The reward system incorporates multiple components in a vectorized form:
\begin{equation}
\mathbf{R}(t) = \text{sum}\bigg(
\begin{bmatrix}
R_{\text{safety}}(\mathbf{d}(t)) \\
R_{\text{efficiency}}(\mathbf{v}(t)) \\
R_{\text{cooperation}}(\mathbf{N}, \gamma)
\end{bmatrix}
\bigg) + \mathbf{A}_{\text{causal}}(t)
\end{equation}
Where $\mathbf{d}(t) \in \mathbb{R}^{N-1}$ is the vector of inter-vehicle distances. $\mathbf{v}(t) \in \mathbb{R}^N$ is the velocity vector of all vehicles. $N$ denotes the total number of vehicles in the system. $\gamma \in (0,1)$ is the discount factor for future rewards. the individual components are defined as safety reward, efficiency reward, and cooperation reward. The safety reward is defined as
\begin{figure}[t]
    \centering
    \includegraphics[width=0.9\linewidth]{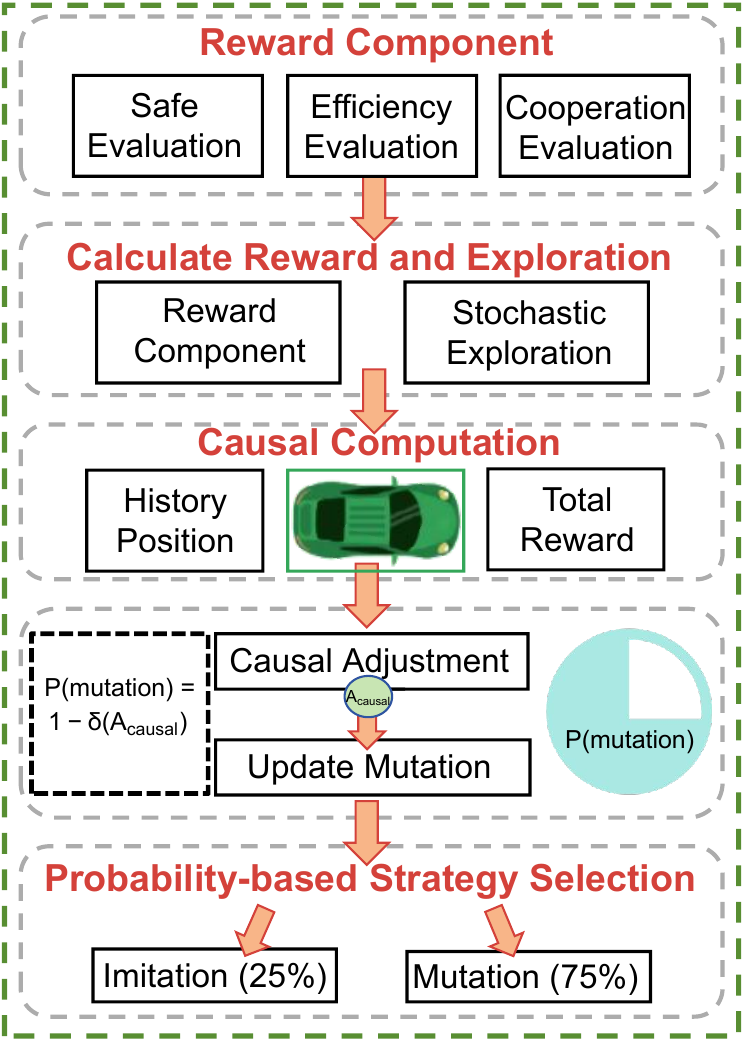}
    \caption{Causal evaluation-based evolutionary game. }
    \label{fig2_interactive}
\end{figure}
\begin{equation}
R_{\text{safety}} = R_{\text{safety\_base}} \cdot \frac{1}{\max(d_{i,i+1}, 1)}
\end{equation}
where $R_{\text{safety\_base}}$ is a constant baseline reward. $d_{i,i+1}$ represents the distance between consecutive vehicles $i$ and $i+1$. Efficiency reward considering individual and average velocities $v_{\text{avg}}$:
\begin{equation}
R_{\text{efficiency}} = R_{\text{efficiency\_base}} \cdot \frac{v_i}{v_{\text{avg}}}, \quad v_{\text{avg}} = \frac{1}{N}\sum_{j=1}^N v_j
\end{equation}
where $v_i$ is the current velocity of vehicle $i$. $R_{\text{efficiency\_base}}$ is a constant baseline reward. Cooperation reward incorporating historical collision statistics:
\begin{equation}
\begin{aligned}
R_{\text{cooperation}} &= (N - 1) \cdot \gamma \cdot R_{\text{cooperation\_base}}(C_{\text{hist}}) \\
&= (N - 1) \cdot \gamma \cdot (-\lambda \cdot \text{rand}([\underline{C}, \overline{C}]))
\end{aligned}
\end{equation}
Where $C_{\text{hist}}$ represents the historical collision statistics. $\lambda > 0$ is a scaling factor for the cooperation penalty. $\gamma \in (0,1)$ remains the temporal discount factor. the collision-dependent base reward is defined as:
\begin{equation}
R_{\text{cooperation\_base}}(C_{\text{hist}}) = -\lambda \cdot \text{rand}([\underline{C}, \overline{C}])
\end{equation}
where $\text{rand}([a,b])$ generates a uniform random number in interval $[a,b]$, With the collision bounds determined by:
\begin{equation}
\begin{aligned}
\underline{C} &= \min_{t' \in [1,t-1]} \sum_{i=1}^{N-1} C_{i,i+1}(t') \\
\overline{C} &= \max_{t' \in [1,t-1]} \sum_{i=1}^{N-1} C_{i,i+1}(t')
\end{aligned}
\end{equation}
Where $C_{i,i+1}(t)$ denotes collision events between consecutive vehicles at time $t$. $[\underline{C}, \overline{C}]$ defines the range of historical collision counts.

The causal adjustment mechanism integrates stochastic exploration with historical influence:
\begin{equation}
A_{\text{causal}} = \text{cm} \cdot \mathcal{N}(0,1) + \overline{C}(t-1)
\end{equation}

Strategy selection probabilities are determined through a sophisticated sigmoid mapping:
\begin{equation}
\begin{aligned}
P_{\text{imitation}} &= \sigma(A_{\text{causal}}) \\
&= \frac{1}{1 + \exp(-\beta(A_{\text{causal}} - \alpha))}
\end{aligned}
\end{equation}

\begin{equation}
P_{\text{mutation}} = 1 - P_{\text{imitation}}
\end{equation}

Where $\sigma(\cdot)$ represents the shifted sigmoid function with parameters $\alpha$ and $\beta$. $\alpha$ controls the inflection point of the sigmoid curve. $\beta$ adjusts the steepness of the probability transition. $\text{cm}$ is the causal mask parameter controlling exploration intensity. $\mathcal{N}(0,1)$ represents the standard normal distribution. $\overline{C}(t-1)$ is the mean historical causal influence from the previous time step. The sigmoid function with its parameters provides a smooth, differentiable mapping from causal adjustment to probability space while maintaining the bounds $P_{\text{imitation}}, P_{\text{mutation}} \in [0,1]$ and ensuring $P_{\text{imitation}} + P_{\text{mutation}} = 1$.

\subsection{Imitation and Mutation in Evolutionary Game Theory}
\begin{algorithm}
\caption{Evolutionary Game Strategy with Imitation and Mutation Controlled by Causal Adjustment}
\begin{algorithmic}[1]
    \For{each vehicle \(i\)}
        \State \textbf{Initialize}: Set initial best speed \(v_i^{\text{best}} = v_i^{\text{current}}\) and best reward \(r_i^{\text{best}} = -\infty\).
        \State \textbf{Calculate Causal Adjustment}:
        \State \quad Compute \(A_{\text{causal}} = \text{cm} \cdot \mathcal{N}(0,1) + \overline{C}(t)\).
        \State \textbf{Determine Probabilities for Imitation and Mutation}:
        \State \quad Compute \(P_{\text{imitation}} = \sigma(A_{\text{causal}})\).
        \State \quad Compute \(P_{\text{mutation}} = 1 - P_{\text{imitation}}\).
        
        \State \textbf{Strategy Selection}:
        \State \quad Generate a random value \(R \in [0, 1]\).
        \If{\(R < P_{\text{imitation}}\)}
            \State \textbf{Imitation Strategy}:
            \State \quad Identify vehicle \(j\) with the highest reward \(r_j\) in the vicinity.
            \State \quad Set trial speed \(v_{\text{trial}} = v_j + \epsilon\), where \(\epsilon\) is a small perturbation.
        \Else
            \State \textbf{Mutation Strategy}:
            \For{trial speed \(v_{\text{trial}}\) from \(v_i^{\text{current}} - 2\) to \(v_i^{\text{current}} + 2\) with step size \(0.1\)}
                \If{\(v_{\text{min}} \leq v_{\text{trial}} \leq v_{\text{max}}\)}
                    \State Update temporary speed and calculate the new position for vehicle \(i\).
                    \State Calculate total temporary reward:
                    \[
                    r_{s,i}^{\text{temp}} = R_{\text{safety}} + R_{\text{efficiency}} + R_{\text{cooperation}} + A_{\text{causal}}
                    \]
                    \If{\(r_{s,i}^{\text{temp}} > r_i^{\text{best}}\)}
                        \State Update \(r_i^{\text{best}} = r_{s,i}^{\text{temp}}\).
                        \State Update \(v_i^{\text{best}} = v_{\text{trial}}\).
                    \EndIf
                \EndIf
            \EndFor
        \EndIf
        \State Set final speed \(v_i = \min(v_i^{\text{best}}, v_{\text{max}})\).
    \EndFor
\end{algorithmic}
\end{algorithm}
The Evolutionary Game Strategy with Imitation and Mutation is designed for each vehicle to dynamically determine its optimal speed through a balance of imitation and mutation strategies. Each decision a vehicle makes is guided by a calculated causal influence, which allows it to learn from past experiences while also exploring new possibilities. Initially, each vehicle starts by setting its current speed $v_i^{\text{current}}$ as the best speed $v_i^{\text{best}}$, a very low value is defined for the best reward:
\begin{equation}
r_i^{\text{best}} = -\infty
\end{equation}
where $-\infty$ means no reward has been obtained yet, and the vehicle is ready to begin its evaluation. This initialization stage is crucial, as it provides the baseline from which each vehicle will begin optimizing its strategy. Next, the $A_{\text{causal}}$ is calculated for each vehicle. This adjustment is an important factor that balances between randomness and learned influence, incorporating past data to inform future behavior. Once the $A_{\text{causal}}$ has been calculated, the vehicle then determines the likelihood of either imitating another vehicle's successful strategy or independently exploring through mutation. The \(P_{\text{imitation}}\) is obtained by passing the \(A_{\text{causal}}\) through a sigmoid function \(\sigma(x)\). 
\subsection{General Algorithm for Multi-Vehicle Interaction}
The vehicle then generates a random value \( R \) between 0 and 1. The decision to imitate another vehicle is made by comparing \( R \) with the imitation probability \( P_{\text{imitation}} \):
\begin{equation}
\text{If } R < P_{\text{imitation}}, \quad \text{then select imitation strategy.}
\end{equation}

Once the imitation strategy is selected, the vehicle identifies another vehicle \( j \) in its vicinity that has the highest reward, denoted as \( r_j \). The trial speed for vehicle \( i \) is then set to be close to the speed of vehicle \( j \), denoted as \( v_j \), but with a small perturbation \( \epsilon \):
\begin{equation}
v_{\text{trial}} = v_j + \epsilon
\end{equation}
where \( \epsilon \sim \mathcal{N}(0, \sigma^2) \) is a small random perturbation drawn from a normal distribution with mean 0 and variance \( \sigma^2 \). This ensures a slight differentiation in the imitated speed to avoid exact replication.

If the random value \( R \) is greater than or equal to \( P_{\text{imitation}} \), the vehicle chooses to perform a mutation:
\begin{equation}
\text{If } R \geq P_{\text{imitation}}, \quad \text{then select mutation strategy.}
\end{equation}

The mutation strategy involves exploring various potential speeds within a specified range. The trial speeds are denoted by \( v_{\text{trial}} \) and are evaluated within the following range:
\begin{equation}
v_{\text{trial}} \in [v_i^{\text{current}} - 2, v_i^{\text{current}} + 2] \quad \text{with increment of } 0.1
\end{equation}
Each \( v_{\text{trial}} \) must satisfy a feasibility condition:
\begin{equation}
v_\text{min} \leq v_{\text{trial}} \leq v_\text{max}
\end{equation}
where $v_\text{min}$ and $v_\text{max}$ are the minimum and max speeds respectively.
\begin{figure*}[t] 
    \centering
    \begin{subfigure}[b]{0.5\textwidth}
        \includegraphics[width=\linewidth]{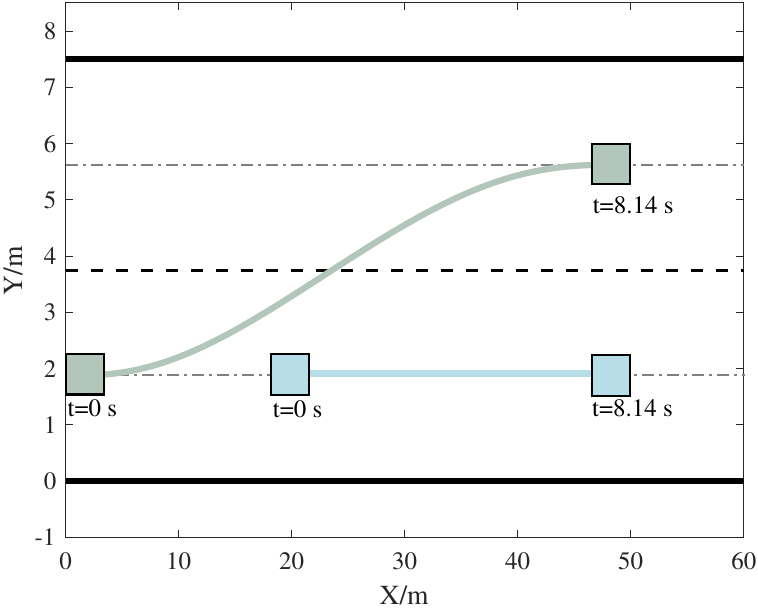}
        \caption{}
        \label{fig11_3d_qulv_sub1}
    \end{subfigure}%
    \begin{subfigure}[b]{0.5\textwidth}
        \includegraphics[width=\linewidth]{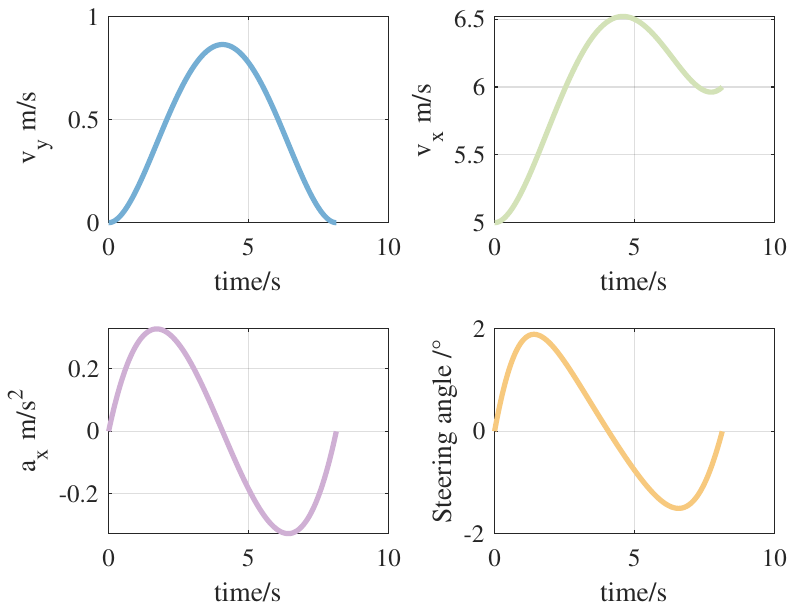}
        \caption{}
        \label{fig11_3d_qulv_sub2}
    \end{subfigure}%
    \vspace{-2mm}
    \caption{Illustration of quintic polynomial path planning. (a) Lane-changing process with another vehicle on the road. (b) Variation of longitudinal speed, lateral speed, longitudinal acceleration, and steering angle.}
    \label{figw}
\end{figure*}  
If the condition is satisfied, the vehicle proceeds by setting its temporary speed to the trial speed and evaluating the corresponding reward:
\begin{equation}
v_i^{\text{temp}} = v_{\text{trial}}
\end{equation}
\begin{algorithm}
\caption{Full Simulation Framework for Multi-Vehicle Interaction}
\begin{algorithmic}[1]
    \State \textbf{Initialize Parameters}:
    \State Set initial positions, speeds, lane positions, rewards, and causal influence metrics for all vehicles.
    \For{each time step \( t \)}
        \For{each vehicle \( i \)}
            \State \textbf{Update Vehicle Kinematics}:
            \State Update position of vehicle \( i \):
            \[
            P_i(t+1) = P_i(t) + v_i(t) \cdot \Delta t
            \]
            \State \textbf{Apply Evolutionary Game Strategy}:
            \State Use imitation or mutation to determine optimal speed \( v_i(t+1) \) 
            \State \textbf{Collision Handling}:
            \If{\( d_{i,i+1} < d_{\text{min}} \)}
                \State Apply collision penalty:
                \begin{equation}
                r_{s,i} = r_{s,i} + P_{\text{collision}}
                \end{equation}
                where $P_{\text{collision}}$ is the collision penalty.
                \State Adjust speed of vehicle \( i \) and preceding vehicle to mitigate collision risk.
            \EndIf
            \State \textbf{Lane Change Decision}:
            \State Generate random value \( R_{\text{lane}} \in [0, 1] \).
            \If{\( R_{\text{lane}} < 0.1 \)}
                \State \textbf{Check Feasibility of Lane Change}:
                \If{Adjacent lane has sufficient spacing}
                    \State Execute lane change maneuver using quintic polynomial trajectory
                \EndIf
            \EndIf
        \EndFor
        \State \textbf{Record Metrics and Update States}:
        \For{each vehicle \( i \)}
            \State Record current rewards \( r_{s,i} \), lane positions, speeds, and causal effectiveness metrics.
        \EndFor
        \State \textbf{Update Historical Records}:
        \State Update positions, speeds, lane change occurrences, and collision counts for all vehicles.
    \EndFor
\end{algorithmic}
\end{algorithm}
After computing the total reward for each trial speed, the vehicle updates its best reward (\(r_i^{\text{best}}\)) and best speed (\(v_i^{\text{best}}\)) if the current trial speed provides a higher reward than previously recorded. This process continues until all trial speeds within the specified range are evaluated. This selection guarantees that each vehicle chooses the most optimal behavior, balancing between safety, efficiency, cooperation, and adaptive learning. 

Fig.~3 illustrates the process of strategy adaptation in a multi-vehicle environment through causal influence and evaluation metrics. Starting at time step \( t = N \), each vehicle utilizes its history Position \( P_1, P_2, \ldots, P_{N-1} \) to perform causal Computation. This process analyzes historical positions and computes a causal influence, which impacts the current strategy. Following causal computation, the causal adjustment at $T=N$, \( A_{\text{causal}, N} \), is applied to influence the vehicles' decisions and strategies at the next time step \( t = N+1 \). Vehicles undergo evaluations on safety, efficiency, and cooperation. These evaluations generate \( R_{\text{safe}}, R_{\text{efficiency}}, R_{\text{cooperation}} \), along with the causal adjustment \( A_{\text{causal}, N-1} \) from previous steps. These elements contribute to the total reward, represented by a diamond shape in the middle of the diagram. This reward guides vehicles' adaptation process through Imitation and Mutation mechanisms. In imitation, vehicles select strategies with the highest reward from past evaluations, while in mutation, vehicles explore extended strategies, enhancing adaptability in dynamic environments. The flow from history position to causal adjustment and reward evaluation, then through imitation or mutation, leads to updated strategies at \( t = N+1 \), promoting optimized and adaptive behavior across all vehicles. The whole process with detailed strategies update is summarized in Algorithm~1

\subsection{Lane-Changing intention using Quintic Polynomial Path Planning}

To ensure smooth and comfortable lane changes while maintaining safety, we implement a quintic polynomial trajectory planning system. The trajectory is defined by a set of boundary conditions in both position and its derivatives:

\begin{equation}
\begin{bmatrix}
1 & 0 & 0 & 0 & 0 & 0 \\
0 & 1 & 0 & 0 & 0 & 0 \\
0 & 0 & 2 & 0 & 0 & 0 \\
1 & T & T^2 & T^3 & T^4 & T^5 \\
0 & 1 & 2T & 3T^2 & 4T^3 & 5T^4 \\
0 & 0 & 2 & 6T & 12T^2 & 20T^3
\end{bmatrix}
\begin{bmatrix}
a_0 \\
a_1 \\
a_2 \\
a_3 \\
a_4 \\
a_5
\end{bmatrix} = 
\begin{bmatrix}
y_0 \\
\dot{y}_0 \\
\ddot{y}_0 \\
y_f \\
\dot{y}_f \\
\ddot{y}_f
\end{bmatrix}
\end{equation}
where \( y_0, \dot{y}_0, \ddot{y}_0 \) represent the initial lateral position, velocity, and acceleration, respectively, and \( y_f, \dot{y}_f, \ddot{y}_f \) represent the corresponding final values at the end of the trajectory. The variable \( T \) denotes the total time duration for the lane change, while \( a_0, a_1, a_2, a_3, a_4, a_5 \) are the coefficients of the quintic polynomial to be solved.

The lateral position \( y(t) \) during the lane change is then given by:
\begin{equation}
y(t) = a_0 + a_1t + a_2t^2 + a_3t^3 + a_4t^4 + a_5t^5
\end{equation}
where \( t \) is the time variable ranging from \( 0 \) to \( T \), and \( y(t) \) represents the lateral position of the vehicle at time \( t \).

The coefficients \( [a_0, a_1, a_2, a_3, a_4, a_5] \) are determined by solving the linear system above, considering the following initial and final conditions:
\[
\text{Initial state: } y_0, \dot{y}_0 = 0, \ddot{y}_0 = 0
\]
\[
\text{Final state: } y_f = y_0 \pm l_{\text{lane}}, \dot{y}_f = 0, \ddot{y}_f = 0
\]
where \( l_{\text{lane}} \) represents the lane width. The sign of \( l_{\text{lane}} \) determines the direction of the lane change: \( + \) for a left lane change and \( - \) for a right lane change. The quintic polynomial ensures a smooth transition with controlled curvature, avoiding abrupt changes in motion and ensuring stability and comfort during the lane change.

Fig.~5(a) illustrates a lane-changing process where a vehicle smoothly transitions from one lane to another using a quintic polynomial trajectory. At the initial time $t=0~\textrm{s}$, the vehicle begins in its original lane with defined position, velocity, and acceleration. Over the lane-changing period $t=8.14~\textrm{s}$, the green vehicle follows a carefully planned trajectory that moves it laterally from the starting lane to the target lane, ensuring a gradual and continuous path. The trajectory allows the vehicle to reach the new lane with zero lateral velocity and acceleration, stabilizing it smoothly in the target lane. This process ensures stability and comfort.

Fig.~5(b) shows the variations of key variables for a green vehicle during a lane change over $8.14~\textrm{s}$. The lateral velocity (\( V_y \)) starts at zero, peaks around 1 m/s at 5 seconds, and returns to zero, reflecting the lateral movement during the maneuver. The longitudinal velocity (\( V_x \)) increases slightly from 6 m/s, reaches a peak just above 6.5 m/s, and then decreases, indicating minor speed adjustments. The longitudinal acceleration (\( a_x \)) oscillates between positive and negative values, supporting the controlled speed variations. Lastly, the steering angle fluctuates around zero, with positive and negative peaks, showing the steering adjustments necessary to execute the lane change smoothly.

\subsection{Collision Detection During Lane Change}
To ensure safety during the lane-changing process, collision detection is performed at each time step. The positions and lane offsets of all vehicles are updated, and the following condition is used to check for potential collisions:
\begin{equation}
\begin{aligned}
| \text{lanes}(i) - \text{lanes}(j) | &< \epsilon, \\
d_{i,i+1} &< d_{\text{safe}}.
\end{aligned}
\end{equation}
where \(\epsilon\) is a small threshold representing lane proximity. $d_{i,i+1}$ is the distance between vehicle $i$ and vehicle $i+1$.\(d_{\text{safe}}\) is the minimum safe distance between vehicles to avoid collisions. If a collision is detected, a penalty, \( P_{\text{collision}} \), is applied. The penalty is directly subtracted from the rewards of both vehicles involved in the collision, and their speeds are reduced to mitigate further collision risks. 

To verify the safety of CEGT, this paper uses time-to-collision (TTC) as an evaluation parameter. The TTC for vehicle pair $(i,j)$ is defined as:
\begin{equation}
\text{TTC}_{ij}(t) = \begin{cases}
\frac{x_j(t) - x_i(t) - l_{\text{veh}}}{v_i(t) - v_j(t)} & \text{if } v_i(t) > v_j(t) \\
\infty & \text{otherwise}
\end{cases}
\end{equation}
Where $l_{\text{veh}}$ is the vehicle length. A collision event is registered when:
\begin{equation}
\text{Collision}_{ij}(t) = 
\begin{cases}
1 & \text{if Condition (1) is satisfied} \\
0 & \text{otherwise}
\end{cases}
\end{equation}
where Condition (1) is defined as:
\begin{equation}
|x_i(t) - x_j(t)| < d_{\text{safe}} \quad \text{and} \quad |y_i(t) - y_j(t)| < w_{\text{lane}}.
\end{equation}

Algorithm~3 provides a comprehensive process for simulating multi-vehicle interactions over multiple time steps. The comprehensive process integrates kinematic updates, evolutionary game-based decision-making, collision handling, lane-changing dynamics, and parameter adaptation. The framework begins with the initialization of all relevant parameters for each vehicle $i$, including initial positions \( P_i \), initial speeds \( v_i \), initial lane assignments \( l_i \), initial rewards \( r_{s,i} \), and \( C_i \) for each vehicle. For each \( t \), each vehicle updates its kinematics by calculating its new position \( P_i(t+1) \) based on the \( v_i(t) \) and time increment \( \Delta t \). The evolutionary game strategy is applied, where each vehicle optimizes its next speed \( v_i(t+1) \) by balancing rewards related to \( R_{\text{safety}} \), \( R_{\text{efficiency}} \), and cooperation \( R_{\text{cooperation}} \). The final decision is also influenced by the causal adjustment \( A_{\text{causal}} \), ensuring adaptive learning through both imitation and mutation.

One significant addition in this framework is collisions handing, explicitly checked after each kinematic update. If the distance between vehicle \( i \) and the preceding vehicle (\( d_{i,i+1} \)) falls below a minimum safe distance \( d_{\text{min}} \), a \( P_{\text{collision}} \) is applied to the rewards \( r_{s,i} \), and the speeds of both vehicles are adjusted. Another important component is the lane-changing during each time step, each vehicle considers lane-changing based on $\epsilon$ and checks for sufficient spacing in the adjacent lane based on collision detection. If feasible, a lane change is executed using a quintic polynomial, ensuring smooth transitions and improving the overall traffic flow.

At each time step, the framework records several metrics, including \( r_{s,i} \), \( l_i \), \( v_i \), and collision counts. These metrics are used for feedback-driven evaluation. Such records ensures that vehicle behavior is continually refined for improved performance over time. These records also provide valuable data for calculating future causal influence \( C_i \), allowing vehicles to adapt strategies based on historical performance.
\begin{figure*}[t]
    \centering
    \includegraphics[width=0.85\linewidth]{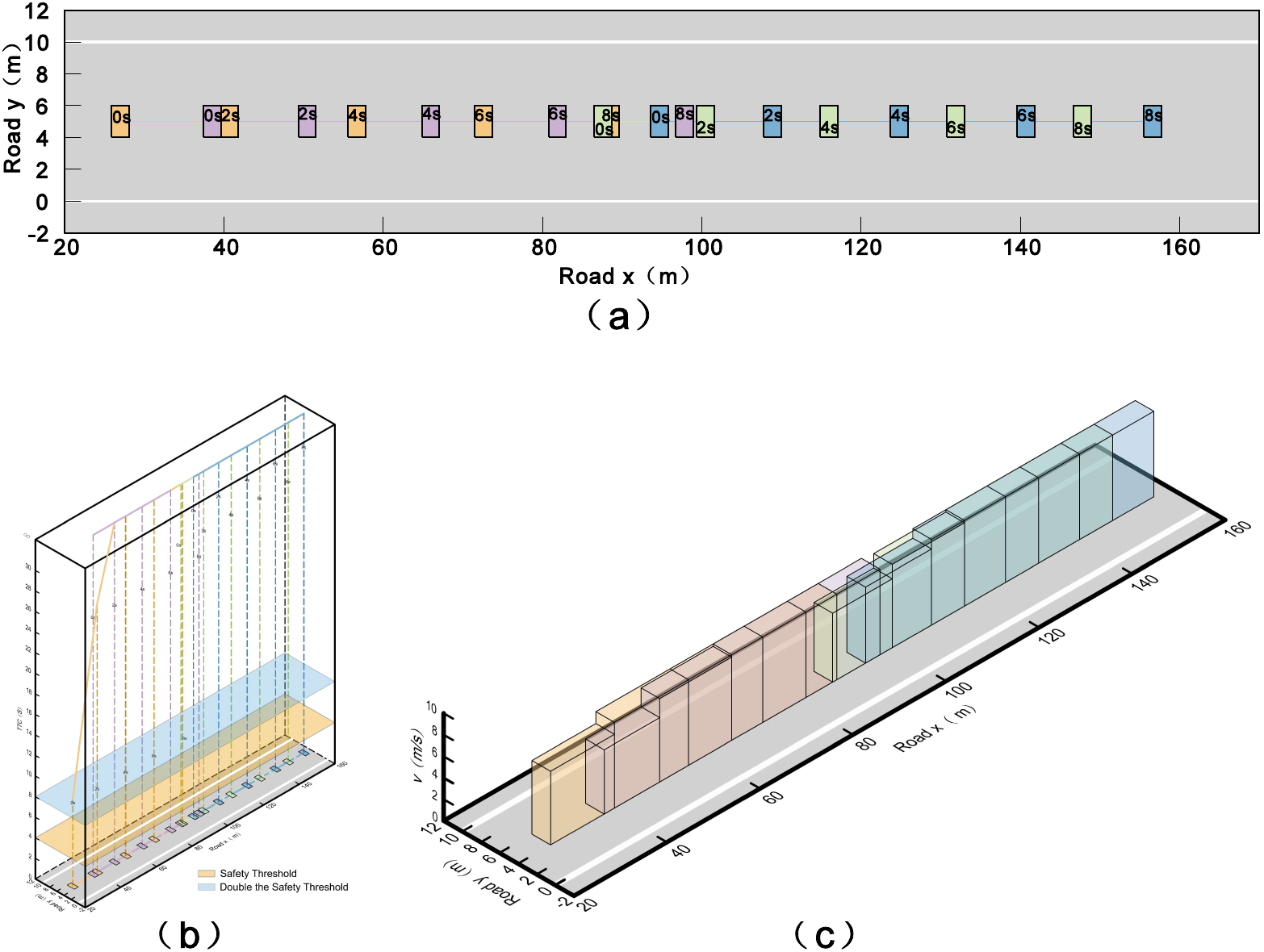}
    \caption{The driving performance of CEGT for Case 1. (a) shows the TTC variations, and (b) shows the speed variations.}
    \label{fig6_interactive}
\end{figure*}


\section{Simulation Results}
\label{sec4}

The simulations are designed to evaluate the safety, efficiency, and cooperativeness of CEGT in two driving scenarios using MATLAB 2024b. To generalize the results, the initial states of autonomous vehicles (AVs) are randomly generated within a predefined range. The safety, efficiency, and cooperativeness of CEGT and three other game-based algorithms have been assessed in terms of the number of collisions, time-to-collision (TTC), average speed, and average reward over various training epochs. The number of collisions achieved by CEGT has been compared and analyzed across 100 random cases. Additionally, the TTC, average speed, and average reward achieved by CEGT have been compared and evaluated over 10 iterations of driving in each simulated scenario. The safe TTC is the 3s-TTC threshold is widely recognized in peer-reviewed literature as a common warning criterion in collision avoidance systems~\cite{fung2003camera,das2019defining}. Considering the potential unknown uncertainties on real-world roads, this paper uses the 4s-TTC as safety threshold.
\begin{figure}[t]
    \centering
    \includegraphics[width=0.9\linewidth]{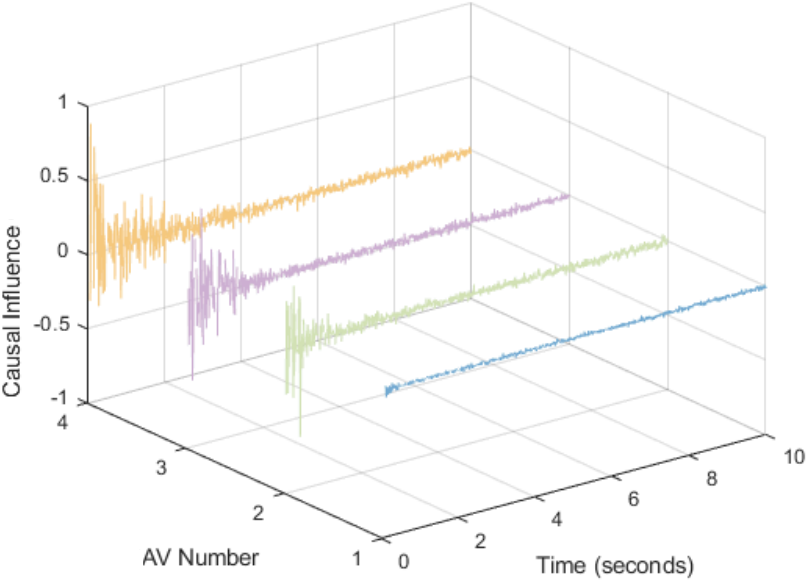}
    \caption{Causal Influence variations for case 1}
    \label{fig7_interactive}
\end{figure}
\begin{figure}[t]
    \centering
    \includegraphics[width=0.8\linewidth]{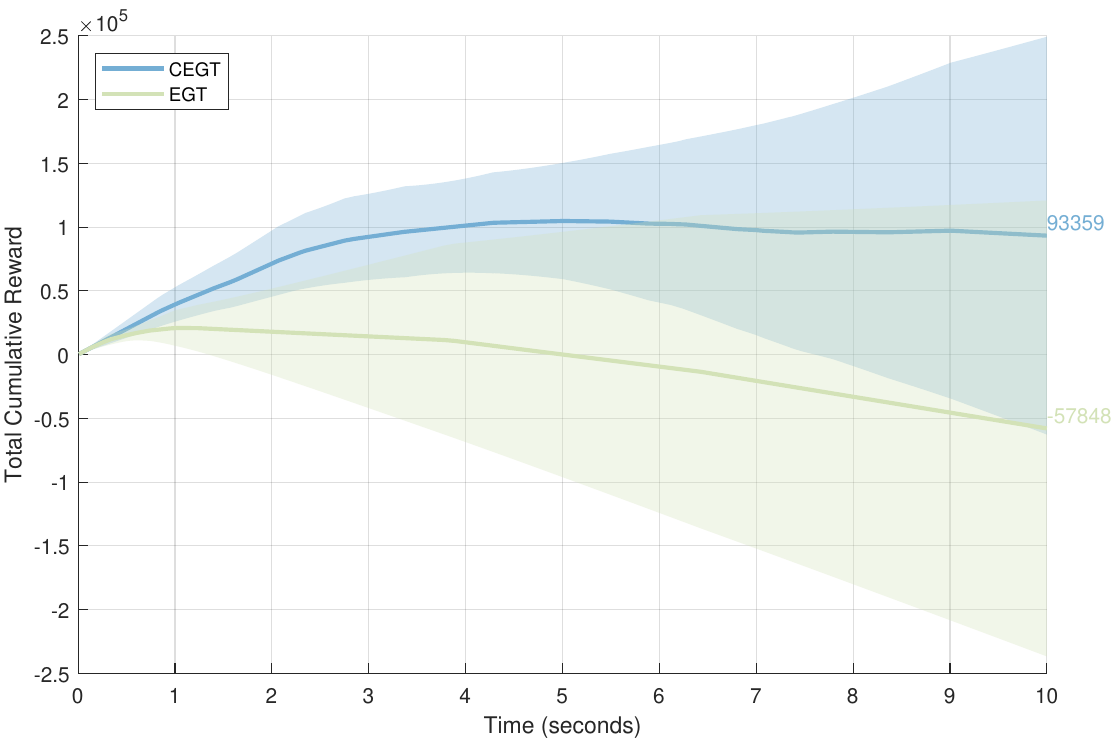}
    \caption{The reward curves of CEGT and EGT across varied initial state settings over 10 runs for Case 1.}
    \label{fig8_interactive}
\end{figure}
\begin{figure}[t]
    \centering
    \includegraphics[width=0.8\linewidth]{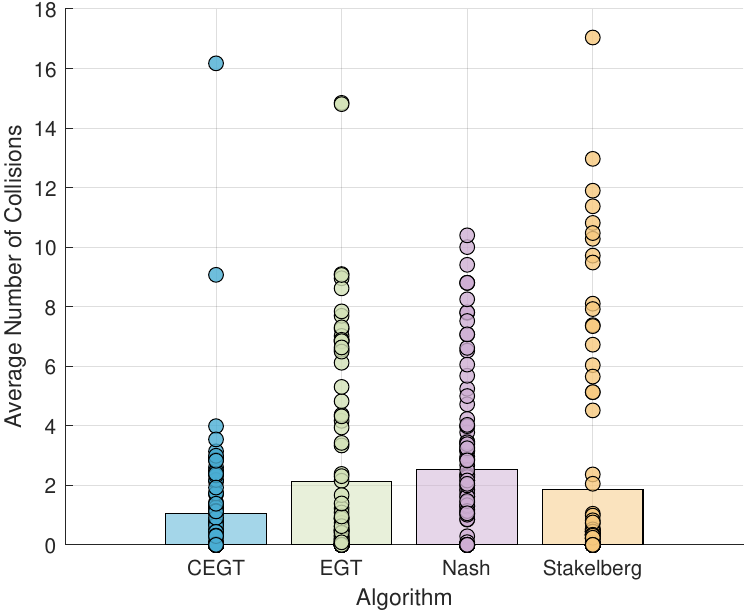}
 \caption{The average number of collisions for CEGT and other benchmark algorithms across varied initial state settings over 100 runs for Case 1.}
    \label{fig9_interactive}
\end{figure}
\begin{figure}[t]
    \centering
    \includegraphics[width=0.8\linewidth]{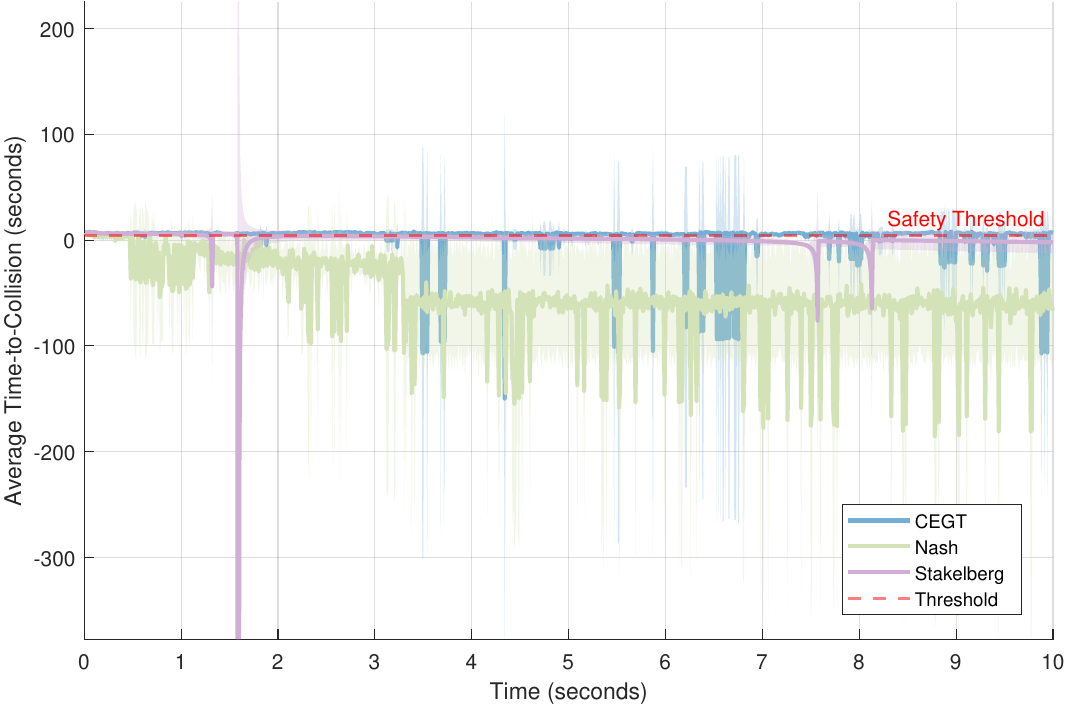}
 \caption{The average TTC curves for all AVs using CEGT and other benchmark algorithms across varied initial state settings over 10 runs for Case 1.}
    \label{fig10_interactive}
\end{figure}
\begin{figure}[t]
    \centering
    \includegraphics[width=0.8\linewidth]{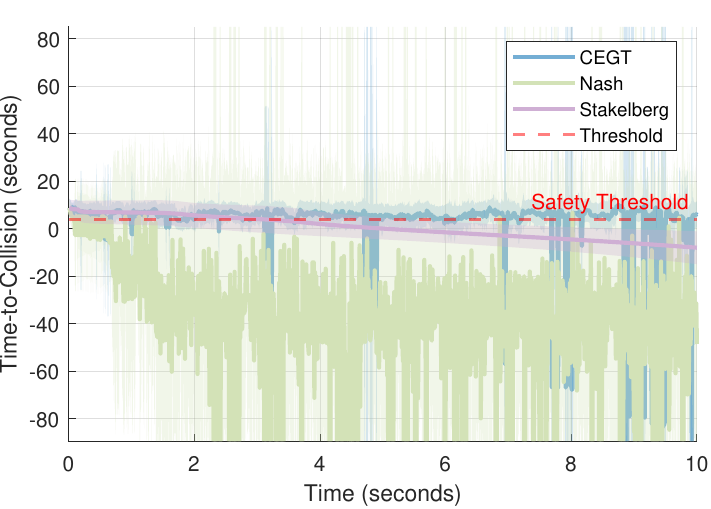}
   \caption{The average TTC curves between AV 1 and AV 2 for CEGT and other benchmark algorithms across varied initial state settings over 10 runs for Case 1.}
    \label{fig11_interactive}
\end{figure}
\begin{figure}[t]
    \centering
    \includegraphics[width=0.8\linewidth]{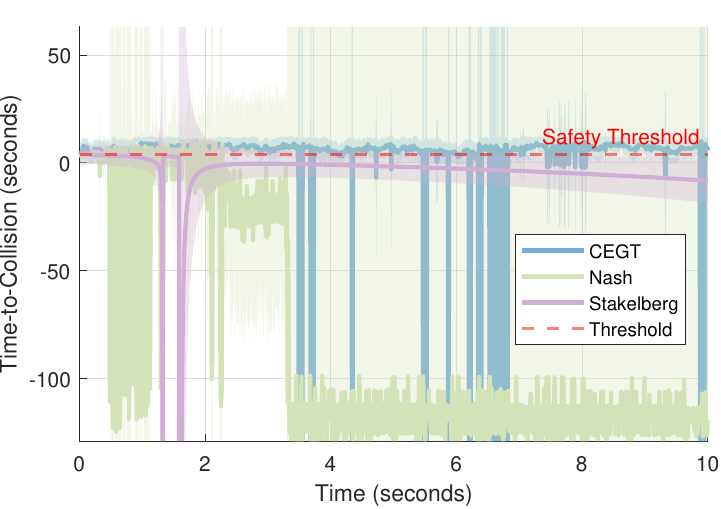}
    \caption{The average TTC curves between AV 3 and AV 4 for CEGT and other benchmark algorithms across varied initial state settings over 10 runs for Case 1.}
    \label{fig12_interactive}
\end{figure}
\begin{figure}[t]
    \centering
    \includegraphics[width=0.8\linewidth]{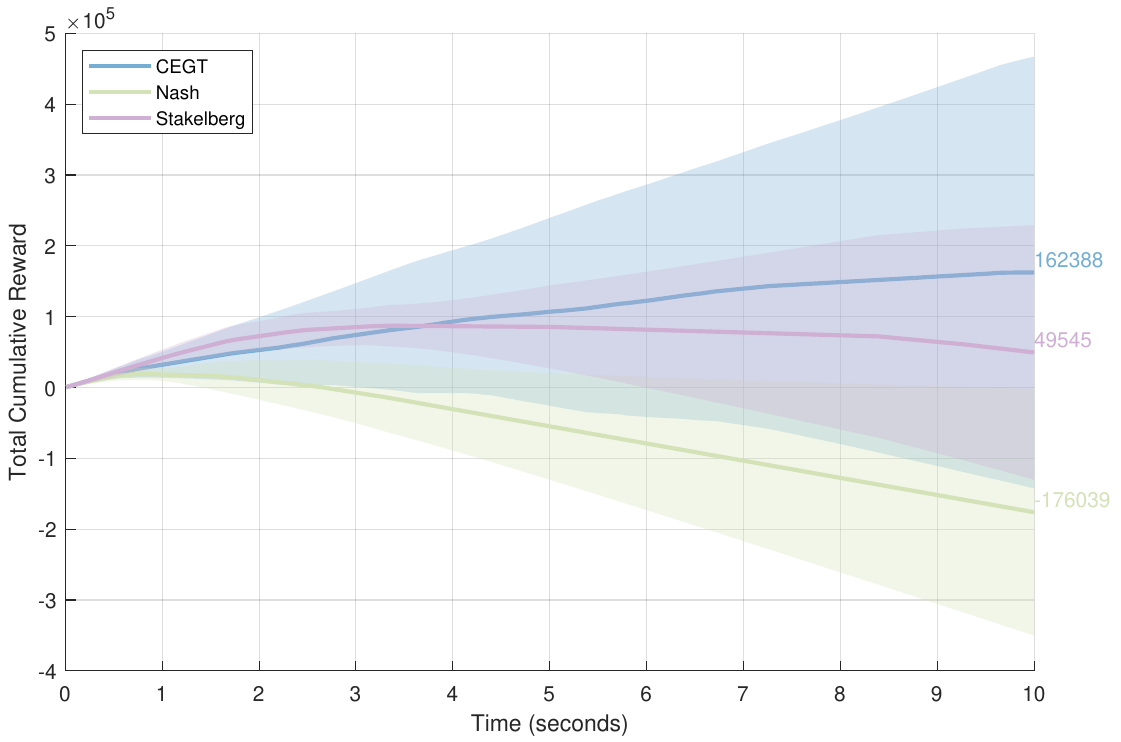}
    \caption{The reward curves of CEGT, Nash and Stackelberg across varied initial state settings over 10 runs for Case 1.}
    \label{fig13_interactive}
\end{figure}
\begin{figure}[t]
    \centering
    \includegraphics[width=0.8\linewidth]{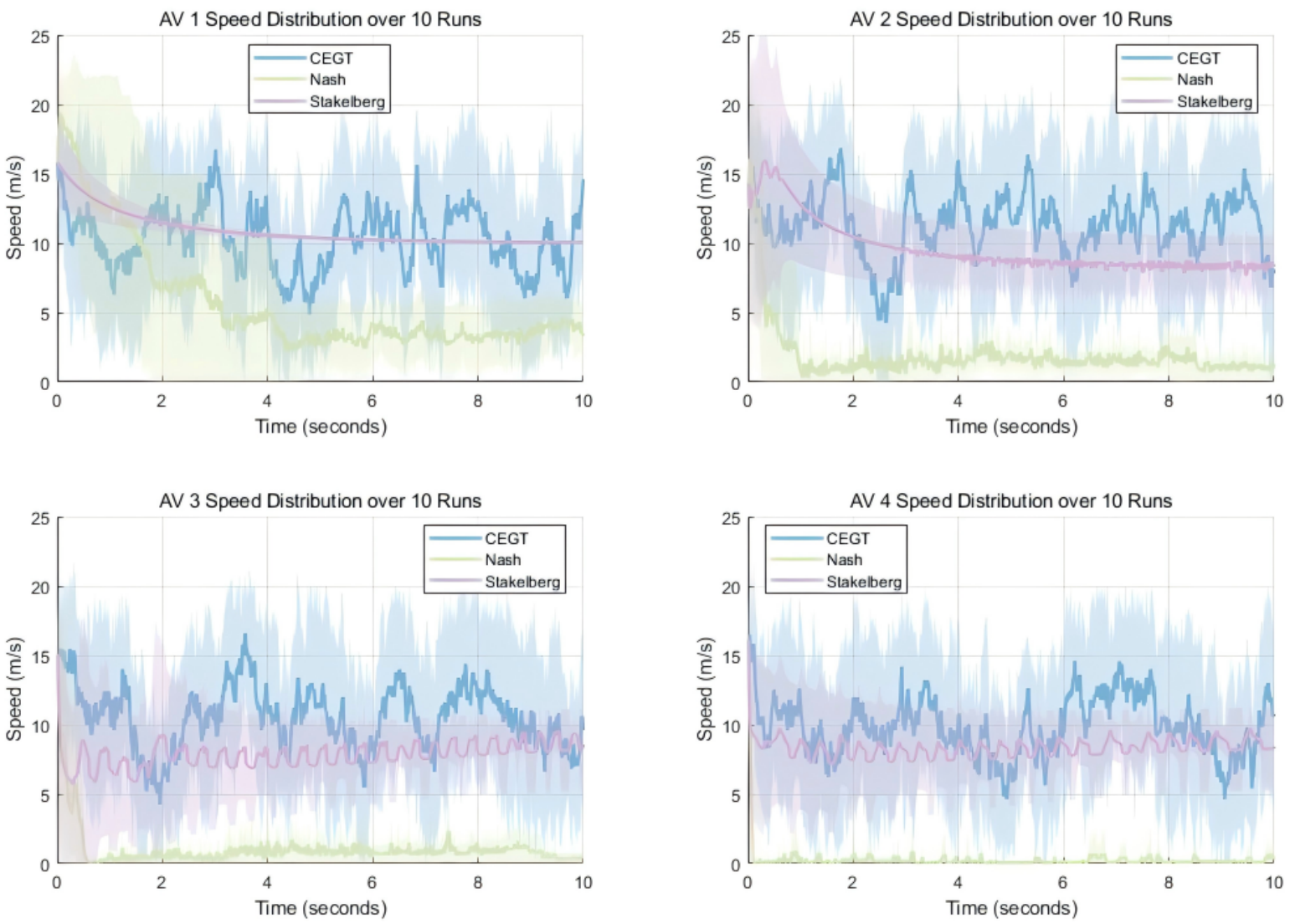}
\caption{The speed curves for each AV using CEGT, Nash, and Stackelberg across varied initial state settings over 10 runs for Case 1.}
    \label{fig14_interactive}
\end{figure}
\begin{figure*}[t]
    \centering
    \includegraphics[width=0.85\linewidth]{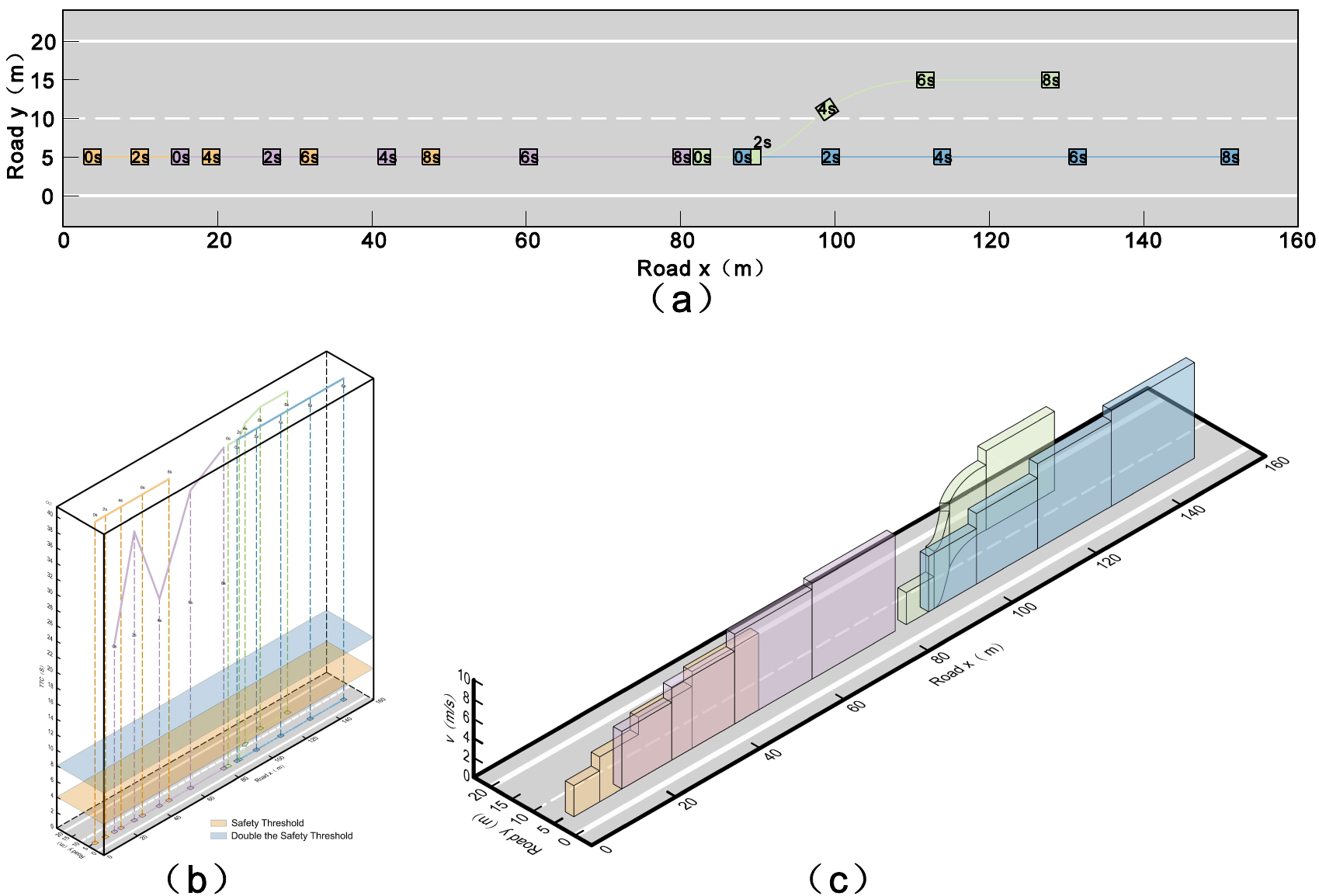}
    \caption{The driving performance of CEGT for Case 2. (a) shows the TTC variations, and (b) shows the speed variations.}
    \label{fig15_interactive}
\end{figure*}
Two scenarios with four AVs are considered to simulate real-world driving conditions and demonstrate the effectiveness of the proposed approach. The experiments include three primary cases:

\textbf{Case 1:} Four AVs are on a single-lane highway. Each AV is initialized with a random speed ranging from $10~\text{m/s}$ to $20~\text{m/s}$ and a random relative distance ranging from $10~\text{m}$ to $20~\text{m}$.

\textbf{Case 2:} Four AVs are on a double-lane highway, with one AV intending to change lanes. Each AV is initialized with a random speed ranging from $10~\text{m/s}$ to $20~\text{m/s}$ and a random relative distance ranging from $10~\text{m}$ to $20~\text{m}$.

\subsection{Simulation Case 1}
Fig.~6 illustrates the decision-making process of EGCT in Case 1. The tasks of the AVs are to maintain stable driving with a low number of collisions and minimize deceleration events to ensure efficiency, all while moving from left to right. Fig.~6(a) shows the road layout with the positions of the AVs plotted at intervals of $2~\text{s}$, represented by green, blue, yellow, and purple rectangles for the four AVs, respectively. No collisions occur during the driving process, as blocks labeled with the same time do not overlap. The relative distances between any two consecutive vehicles do not significantly increase, indicating overall driving efficiency.

Fig.~6(b) depicts the TTC of each AV throughout the driving process. At each time point, the TTC of every AV is significantly larger than double the safety threshold, indicating extremely high driving safety. Fig.~6(c) illustrates the speed variations of the AVs. The speed of each AV either steadily increases or remains relatively consistent at high levels, reflecting driving efficiency and a cooperative driving pattern, as no deceleration occurs and similar speeds are maintained.

Fig.~7 illustrates the causal influence of the four AVs over time. The numbering of the AVs corresponds to their initial longitudinal positions; for example, AV 1 represents the vehicle farthest from the zero point. Initially, all AVs exhibit significant fluctuations, indicating an adaptation phase during which they adjust to interactions within their environment. Over time, these influences stabilize at different levels, reflecting heterogeneous adaptation among the vehicles. Eventually, all AVs achieve a positive causal influence, effectively contributing to the system, though the times to stabilization vary. AV 1 stabilizes around $1~\text{s}$, AV 2 and AV 3 stabilize around $2~\text{s}$, and AV 4 stabilizes around $3~\text{s}$. This varying trend underscores the importance of adaptive mechanisms in EGT, enabling each AV to adjust independently based on its experiences. AVs with positive causal influence enhance system efficiency and stability, while those with negative influences may require further optimization.

Fig.~8 compares the total cumulative reward of CEGT and EGT over time, illustrating the effectiveness of the causal evaluation mechanism. CEGT consistently outperforms EGT, achieving a significantly higher cumulative reward (\textasciitilde93,359) compared to EGT (\textasciitilde57,848) at $t=10~\text{s}$. This demonstrates superior system efficiency and effectiveness. CEGT also exhibits a faster convergence, with a steep initial increase in rewards, highlighting its efficient adaptation phase, whereas EGT adapts more slowly. Moreover, CEGT stabilizes at higher reward levels with narrower confidence bounds, reflecting consistent and reliable performance. In contrast, EGT shows greater variability, indicating less predictable outcomes. Overall, the causal evaluation mechanism employed by CEGT enhances system efficiency, adaptability, and stability, making it a more effective approach for cooperative decision-making.

Fig.~9 compares the average number of collisions for four algorithms: CEGT, EGT, Nash, and Stackelberg. CEGT demonstrates a significant advantage with an average collision count of approximately $1.5$, the lowest among all algorithms, compared to EGT ($2.8$), Nash ($4.5$), and Stackelberg ($3.6$). The data points for CEGT are tightly concentrated around the mean, reflecting consistent and reliable results, whereas Nash and Stackelberg show wider distributions with extreme outliers, reaching collision counts as high as $16$ and $18$, respectively. This stark contrast highlights the inefficiency of the alternative methods in ensuring safety. CEGT's ability to minimize collisions and maintain stability across scenarios underscores its effectiveness in cooperative and adaptive decision-making.

Fig.~10 illustrates the average TTC for CEGT, Nash, and Stackelberg over time, highlighting how well each algorithm balances safety and efficiency. CEGT maintains TTC values consistently above the safety threshold, ensuring a low risk of collisions throughout the simulation. In contrast, Nash and Stackelberg frequently fall below the safety threshold, particularly around $t=2~\text{s}$, where their TTC values drop as low as $-300~\text{s}$, indicating a high risk of collisions. CEGT also avoids prolonged periods of negative TTC, which occur when the rear AV's speed is slower than the lead AV's, resulting in inefficiencies. Its TTC values remain mostly positive, demonstrating that it maintains efficient driving speeds. In comparison, Nash and Stackelberg often exhibit negative TTC values or positive values below the safety threshold for extended periods. For instance, at $t=5~\text{s}$, Nash has an average TTC value below $-100~\text{s}$, indicating poor speed control. Additionally, CEGT demonstrates stable and consistent performance, with its TTC values generally staying above the safety threshold without significant fluctuations. In contrast, Nash and Stackelberg show far less stability, with frequent and severe drops in TTC, making their performance both unsafe and inefficient.

Fig.~11 illustrates the average TTC between AV 1 and AV 2 for the CEGT, Nash, and Stackelberg algorithms. CEGT demonstrates superior performance by maintaining TTC values consistently close to or above the safety threshold for most of the simulation. CEGT maintains TTC values above safety threshold on average, ensuring low collision risk. In contrast, Nash and Stackelberg frequently exhibit TTC values below the safety threshold, with Nash dropping to as low as $-80~\text{s}$, particularly around $t=2~\text{s}$, indicating high collision risks and inefficiency. CEGT also shows stable behavior, with fewer fluctuations compared to the other algorithms. This stability highlights its ability to manage interactions between AV 1 and AV 2 effectively. On the other hand, Nash and Stackelberg experience significant volatility, with frequent drops in TTC below $0~\text{s}$, reflecting poor speed control and increased risk of unsafe situations.

Fig.~12 illustrates the average TTC between AV 3 and AV 4 for the CEGT, Nash, and Stackelberg algorithms. CEGT demonstrates superior performance by maintaining TTC values consistently above the safety threshold for most of the simulation, ensuring a low risk of collisions. In contrast, Nash and Stackelberg frequently fall below the safety threshold for extended periods, indicating higher collision risks. CEGT also avoids prolonged periods of negative TTC values, which occur when AV 3's speed is slower than AV 4's, leading to inefficiencies. Moreover, CEGT's TTC values remain mostly positive, highlighting effective and efficient speed management. In comparison, Nash and Stackelberg exhibit significant volatility with frequent negative TTC values, reflecting poor coordination and inefficiencies.

Fig.~13 The illustrates the total cumulative reward over time for the CEGT, Nash, and Stackelberg algorithms. CEGT clearly outperforms the other algorithms by achieving the highest cumulative reward throughout the simulation. At $t = 10~\text{s}$, CEGT reaches a reward of approximately $162,388$, significantly higher than Nash around $49,545$ and Stackelberg about $-176,039$. This demonstrates CEGT’s superior ability to balance safety, efficiency, and cooperativeness. CEGT also shows a steeper growth rate in cumulative reward, reflecting its effective decision-making and efficient coordination among vehicles. In contrast, Nash has a much slower growth, while Stackelberg consistently exhibits a decline in cumulative reward, reflecting inefficient performance and frequent unsafe behaviors. Additionally, CEGT maintains a smaller variability range compared to Nash and Stackelberg, as shown by the narrower shaded region. This indicates more stable performance across different initial settings. In comparison, Nash and Stackelberg exhibit larger variability, highlighting their instability and less reliable outcomes.

Fig.~14 illustrates the speed distributions with variations over time for four AVs using the CEGT, Nash, and Stackelberg algorithms across 10 runs. CEGT demonstrates clear advantages in maintaining higher and more stable speeds compared to Nash and Stackelberg. For AV 1, CEGT has higher average speed than Nash and Stackelberg at most time points after $2~\text{s}$.  For AV 2 to AV 4, CEGT achieves an average speed of approximately $10~\text{m/s}$ to $15~\text{m/s}$ throughout the simulation, indicating efficient driving behavior. In contrast, Nash and Stackelberg show significantly lower average speeds, with Stackelberg often around $5~\text{m/s}$ to $10~\text{m/s}$, and Nash below $5~\text{m/s}$, highlighting inefficiencies. 

\subsection{Simulation Case 2}
Fig.~15 illustrates the decision-making process of EGCT in Case 2. The tasks of the AVs are to maintain stable driving with a low number of collisions and minimize deceleration events to ensure efficiency during both stable driving and lane-changing. Fig.~15(a) shows the road layout with the positions of the AVs plotted at intervals of $2~\text{s}$, represented by green, blue, yellow, and purple rectangles for the four AVs, respectively. No collisions occur during either stable driving or lane-changing, as blocks labeled with the same time do not overlap. The relative distances between any two AVs remain consistent, indicating overall driving efficiency.

Fig.~15(b) depicts the TTC of each AV during both stable driving and lane-changing. The purple AV experiences fluctuating TTC due to its lead AV's lane-changing intention but always maintains values larger than double the safety threshold. At each time point, the TTC of the other AVs is also significantly larger than double the safety threshold, indicating extremely high driving safety. Fig.~15(c) illustrates the speed variations of the AVs. The speed of each AV consistently increases, reflecting driving efficiency and a cooperative driving pattern, as no deceleration occurs and similar speeds are maintained at the final time.

Fig.~16 illustrates the causal influence of the four AVs over time, showcasing their adaptive behavior and contributions to the system. The causal influence values are plotted for each AV, represented by different colors. Initially, all AVs exhibit fluctuations, reflecting an adaptation phase where the AVs adjust to the interactions in their environment. Over time, the causal influence stabilizes at different levels for each AV, highlighting heterogeneous adaptation. Positive causal influence values indicate contributions because they reflect how an AV's behavior positively impacts the overall system's reward. When an AV has a positive causal influence, it implies that its actions are cooperative, reduce collision risks, and maintain efficient traffic flow. AV 1 starts with a low causal influence but gradually increases over time, reaching a positive influence near $t = 2~\text{s}$. AV 2 maintains a more stable but lower positive influence throughout the simulation, indicating moderate contributions to the system. AV 3 shows noticeable fluctuations during the adaptation phase but achieves a stable and increasing positive influence by the end. AV 4 starts with the highest fluctuation and influence initially but stabilizes at a consistently high positive causal influence, demonstrating its significant contribution to the system. The positive causal influence of all AVs by the end of the simulation reflects their cooperative behavior, with AV 4 contributing the most and AV 2 contributing moderately. 
\begin{figure}[t]
    \centering
    \includegraphics[width=0.8\linewidth]{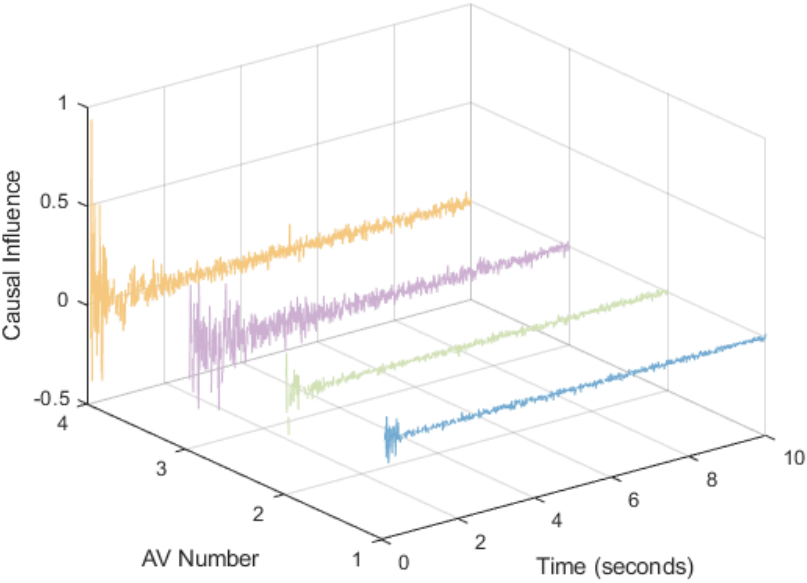}
    \caption{Causal Influence variations for case 2.}
    \label{fig16_interactive}
\end{figure}
\begin{figure}[t]
    \centering
    \includegraphics[width=0.8\linewidth]{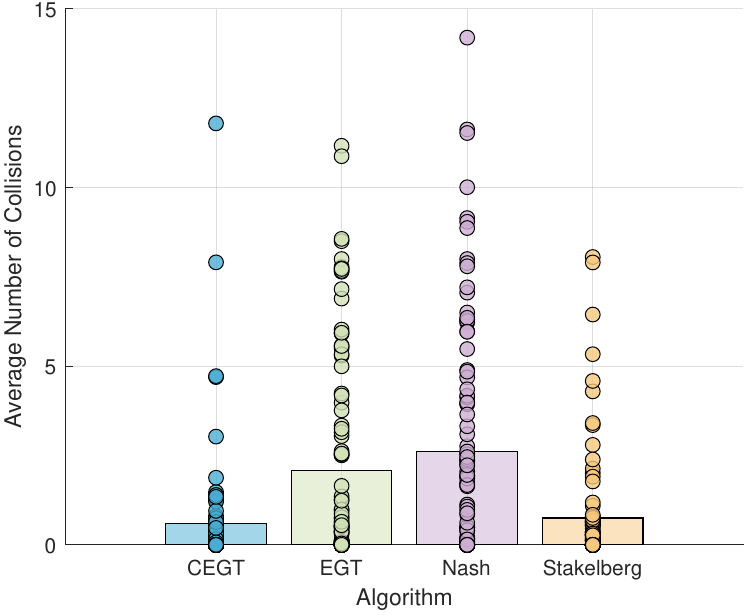}
    \caption{The average number of collisions for CEGT and other benchmark algorithms across varied initial state settings over 100 runs for Case 2.}
    \label{fig17_interactive}
\end{figure}
Fig.~17 demonstrates the average number of collisions for four algorithms: CEGT, EGT, Nash, and Stackelberg. CEGT demonstrates a significant advantage with an average collision count of approximately $0.4$, the lowest among all algorithms, compared to EGT around $2.5$, Nash around $3$, and Stackelberg around $0.7$. The data points for CEGT are tightly concentrated around the mean, reflecting consistent and reliable results, whereas Nash and Stackelberg show wider distributions with extreme outliers between $5$ to $10$. This stark contrast highlights the inefficiency of the alternative methods in ensuring safety. CEGT's ability to minimize collisions and maintain stability across scenarios underscores its effectiveness in cooperative and adaptive decision-making.

Fig.~18 illustrates the cumulative reward profiles over time for four AVs using the CEGT, Nash, and Stackelberg algorithms. For AV 1, CEGT shows a steady growth in cumulative reward, reaching approximately $0.7 \times 10^5$ at $t=10~\text{s}$. This lower reward compared to other AVs is due to AV 1 prioritizing maintaining a similar distance to other AVs, thereby sacrificing its self-reward for the overall system's benefit. For AV 2, the CEGT reward curve is similar to Stackelberg; however, CEGT exhibits narrower variation, suggesting greater stability and adaptability to different initial settings of AVs. Similar trends are observed for AVs 3 and 4, where CEGT achieves higher rewards than the other algorithms. Although Stackelberg also presents a consistent upward trend, its wide variability indicates poor capability to adapt to random initial settings of AVs and maintain reliable performance.
\begin{figure}[t]
    \centering
    \includegraphics[width=0.8\linewidth]{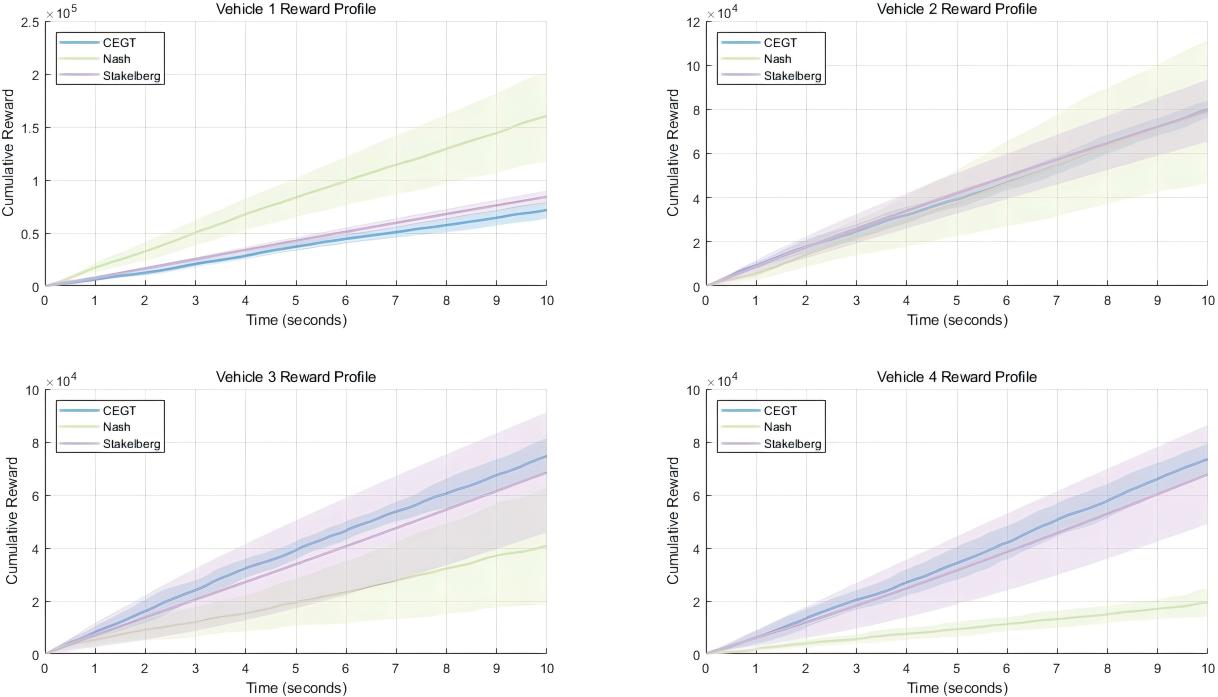}
    \caption{The reward curves of CEGT, Nash and Stackelberg across varied initial state settings over 10 runs for Case 2.}
    \label{fig18_interactive}
\end{figure}
\begin{figure}[t]
    \centering
    \includegraphics[width=0.8\linewidth]{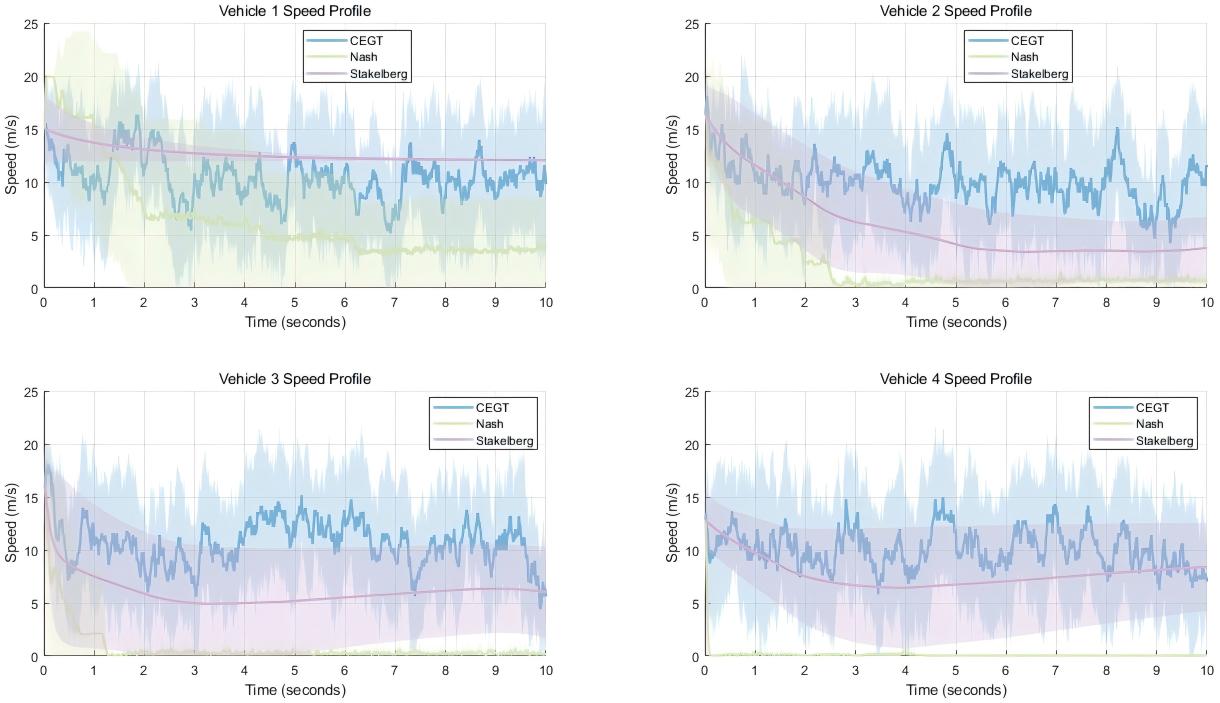}
   \caption{The speed curves for each AV using CEGT, Nash, and Stackelberg across varied initial state settings, with a collision penalty of $-100$, over 10 runs for Case 2.}
    \label{fig19_interactive}
\end{figure}
\begin{figure}[t]
    \centering
    \includegraphics[width=0.8\linewidth]{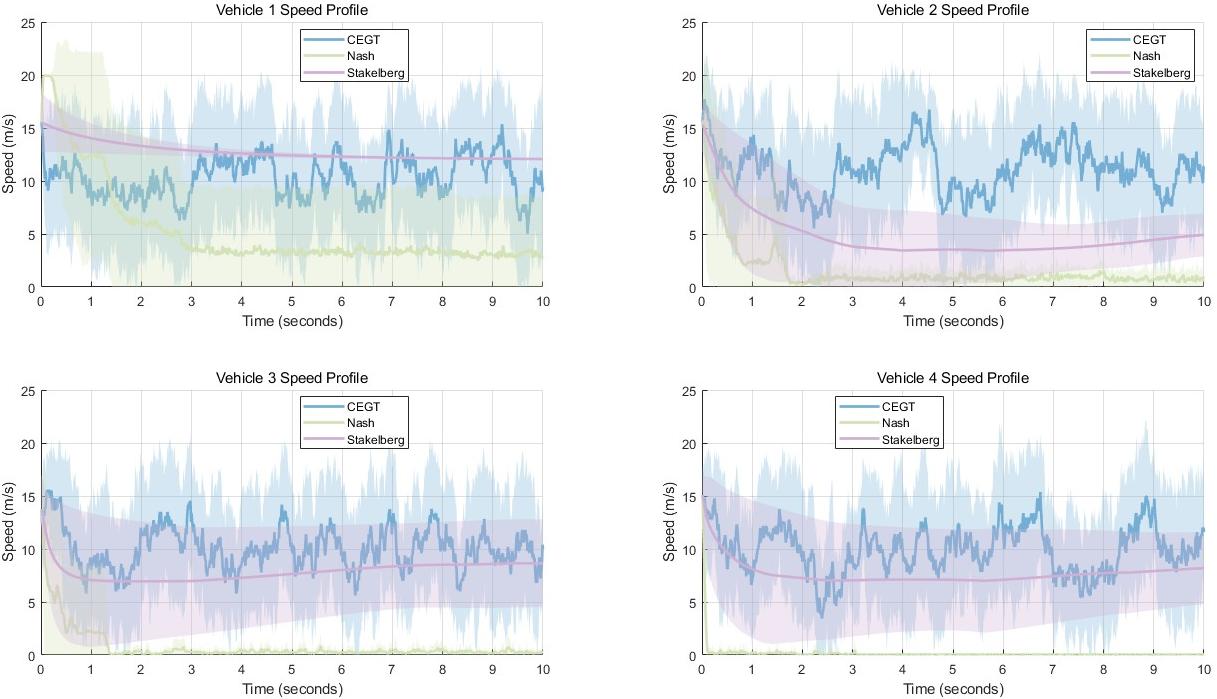}
\caption{The speed curves for each AV using CEGT, Nash, and Stackelberg across varied initial state settings, with a collision penalty of $-200$, over 10 runs for Case 2.}
    \label{fig20_interactive}
\end{figure}
\begin{figure}[h]
    \centering
    \includegraphics[width=0.8\linewidth]{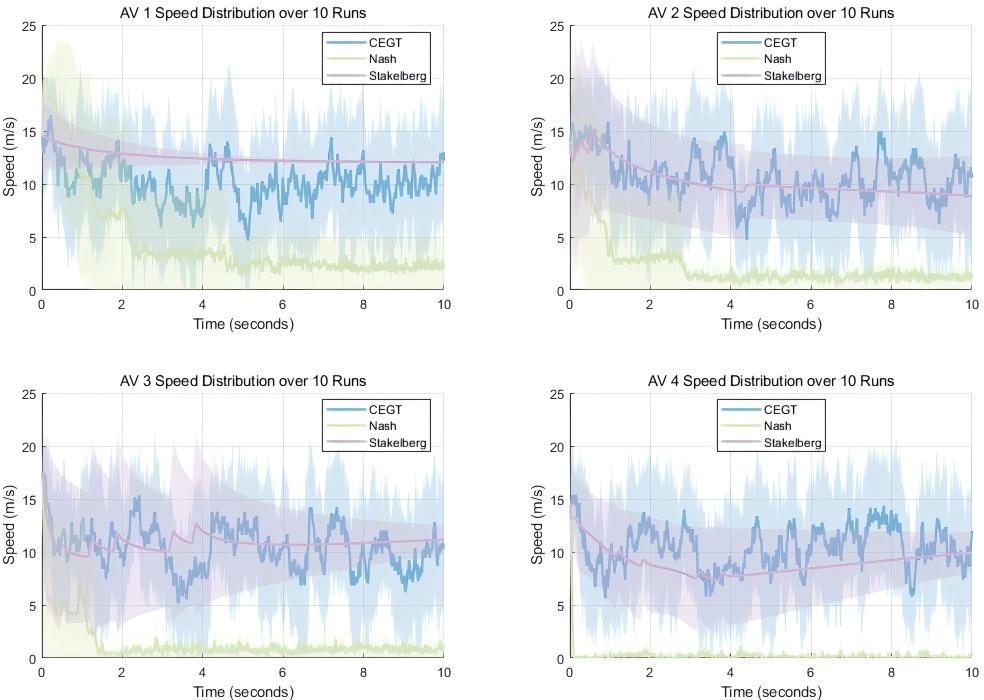}
\caption{The speed curves for each AV using CEGT, Nash, and Stackelberg across varied initial state settings, with a collision penalty of $-300$, over 10 runs for Case 2.}
    \label{fig121_interactive}
\end{figure}
To verify the generalization of CEGT, different algorithm settings are also tested. Fig.~19 shows the speed distributions with variations over time for four AVs using the CEGT, Nash, and Stackelberg algorithms across 10 runs, with a collision penalty of -100. CEGT demonstrates clear advantages in maintaining higher and more stable speeds compared to Nash and Stackelberg. For AV 1, CEGT achieves a lower average speed than Nash and Stackelberg at most time points due to its cooperation-oriented decision-making. This sacrifice by AV 1 allows other AVs to achieve significantly higher speeds. For AVs 2 to 4, CEGT maintains an average speed of approximately $10~\text{m/s}$ to $15~\text{m/s}$ throughout the simulation, reflecting efficient driving behavior. In contrast, Nash and Stackelberg show significantly lower average speeds, with Stackelberg remaining below $10~\text{m/s}$ and Nash often below $5~\text{m/s}$, highlighting inefficiencies.

Fig.~20 shows the speed distributions with variations over time for four AVs using the CEGT, Nash, and Stackelberg algorithms across 10 runs, with a collision penalty of -200. For AV 1, CEGT displays a lower average speed compared to Nash and Stackelberg at most time points, as it prioritizes cooperation over individual performance. This trade-off enables other AVs to achieve substantially higher speeds. For AVs 2 to 4, CEGT consistently maintains an average speed ranging from $10~\text{m/s}$ to $15~\text{m/s}$ throughout the simulation, showcasing its ability to ensure efficient driving behavior. In comparison, Nash and Stackelberg exhibit significantly reduced average speeds, with Stackelberg staying below $10~\text{m/s}$ and Nash frequently falling under $5~\text{m/s}$, underscoring their lack of efficiency.

Fig.~21 shows the speed distributions with variations over time for four AVs using the CEGT, Nash, and Stackelberg algorithms across 10 runs, with a collision penalty of -300. The AV 1's speed patterns show distinct characteristics across the three strategies. The CEGT approach maintains speeds between $8\text{-}12~\text{m/s}$ with moderate fluctuations, while demonstrating higher stability compared to other methods. The Nash equilibrium exhibits a sharp decline in speed after the initial period, settling at considerably low speeds around $2\text{-}3~\text{m/s}$. The Stackelberg approach achieves intermediate performance, maintaining speeds around $10\text{-}12~\text{m/s}$ with relatively consistent behavior. For AVs 2 through 4, similar patterns emerge but with notable differences. CEGT consistently demonstrates superior performance, maintaining speeds between $10\text{-}15~\text{m/s}$ across all vehicles, though with visible oscillations in the velocity profiles. The Nash equilibrium consistently performs poorly across all AVs, with speeds dropping dramatically to near-zero levels after the initial phase. The Stackelberg strategy shows intermediate performance but exhibits greater variance as indicated by the wider shaded regions. The shaded areas indicate that CEGT maintains relatively consistent behavior across multiple simulations, while Nash and Stackelberg strategies show higher variability. This suggests that CEGT not only achieves better average performance but also provides more reliable and predictable vehicle behavior.

\section{Conclusion}
\label{sec5}
    This paper presents a novel EGT-based framework for AV interactions that addresses the challenges faced by modern transportation systems in ensuring road safety and optimizing AV interactions. By introducing a CEGT to optimize the evolutionary rate, the proposed approach balances mutation and evolution by learning from historical interactions. The framework effectively overcomes the limitations of traditional rule-based, optimization-based, and game-theoretic methods, such as lack of adaptability, high computational load, low cooperativeness, and reliance on leader-follower-based decision-making. Extensive simulations demonstrate that the proposed CEGT outperforms EGT and popular benchmark algorithms, including Nash and Stackelberg games, in diverse scenarios and parameter settings. Simulation results highlight lower collision rates, improved safety distances, higher speeds, and overall superior performance in cooperative and decentralized settings. The adaptability and efficiency of the proposed framework ensure its applicability to complex and dynamic driving environments, addressing the limitations of centralized control systems and balancing passenger demands with traffic efficiency. Future research will focus on two key aspects: 1) enhancing the scalability of the framework to handle large-scale AV interactions across varying traffic densities, and 2) incorporating advanced predictive models to further optimize the adaptability and safety of AV interactions under highly dynamic scenarios.

\bibliographystyle{IEEEtran}
\bibliography{IEEEabrv,zq_lib}

@ARTICLE{10048483,
  author={Lee, Taeheon and Choi, Seibum B.},
  journal={IEEE Transactions on Vehicular Technology}, 
  title={Real-Time Optimization of Gear Shift Trajectories Using Quadratic Programming for Electric Vehicles With Dual Clutch Transmission}, 
  year={2023},
  volume={72},
  number={7},
  pages={8647-8660},
  doi={10.1109/TVT.2023.3246179}}

@ARTICLE{10285563,
  author={Kojchev, Stefan and Hult, Robert and others},
  journal={IEEE Transactions on Intelligent Transportation Systems}, 
  title={A Two-Stage MIQP-Based Optimization Approach for Coordinating Automated Electric Vehicles in Confined Sites}, 
  year={2024},
  volume={25},
  number={2},
  pages={2061-2075},
  doi={10.1109/TITS.2023.3320168}}

@article{hang2021decision,
  title={Decision making of connected automated vehicles at an unsignalized roundabout considering personalized driving behaviours},
  author={Hang, Peng and Huang, Chao and Hu, Zhongxu and Xing, Yang and Lv, Chen},
  journal={IEEE Trans. Veh. Technol.},
  volume={70},
  number={5},
  pages={4051--4064},
  year={2021},
  publisher={IEEE}
}

@article{BADUE2021113816,
title = {Self-driving cars: A survey},
journal = {Expert Systems with Applications},
volume = {165},
pages = {113816},
year = {2021},
issn = {0957-4174},
doi = {https://doi.org/10.1016/j.eswa.2020.113816},

author = {Claudine Badue and Rânik Guidolini and Raphael Vivacqua Carneiro and Pedro Azevedo and Vinicius B. Cardoso and Avelino Forechi and Luan Jesus and Rodrigo Berriel and Thiago M. Paixão and Filipe Mutz and Lucas {de Paula Veronese} and Thiago Oliveira-Santos and Alberto F. {De Souza}}
}

@inproceedings{inproceedings,
author = {Pérez, Joshué and Milanes, Vicente and Tere, De and Vlacic, Ljubo},
year = {2011},
month = {08},
pages = {},
title = {Autonomous driving manoeuvres in urban road traffic environment: A study on roundabouts},
volume = {18},
isbn = {9783902661937},
journal = {IFAC Proceedings Volumes (IFAC-PapersOnline)},
doi = {10.3182/20110828-6-IT-1002.00423}
}

@article{ignatious2022overview,
  title={An overview of sensors in Autonomous Vehicles},
  author={Ignatious, Henry Alexander and Khan, Manzoor and others},
  journal={Procedia Computer Science},
  volume={198},
  pages={736--741},
  year={2022},
  publisher={Elsevier}
}

@article{ding2019multivehicle,
  title={Multivehicle coordinated lane change strategy in the roundabout under internet of vehicles based on game theory and cognitive computing},
  author={Ding, Nan and Meng, Xianghua and Xia, Weiguo and Wu, Di and Xu, Li and Chen, Bingcai},
  journal={IEEE Transactions on Industrial Informatics},
  volume={16},
  number={8},
  pages={5435--5443},
  year={2019},
  publisher={IEEE}
}

@article{hang2020integrated,
  title={An integrated framework of decision making and motion planning for autonomous vehicles considering social behaviors},
  author={Hang, Peng and Lv, Chen and Huang, Chao and Cai, Jiacheng and Hu, Zhongxu and Xing, Yang},
  journal={IEEE transactions on vehicular technology},
  volume={69},
  number={12},
  pages={14458--14469},
  year={2020},
  publisher={IEEE}
}

@article{deluka2018introduction,
  title={Introduction of Autonomous Vehicles: Roundabouts design and safety performance evaluation},
  author={Deluka Tiblja{\v{s}}, Aleksandra and Giuffr{\`e}, Tullio and Surdonja, Sanja and Trubia, Salvatore},
  journal={Sustainability},
  volume={10},
  number={4},
  pages={1060},
  year={2018},
  publisher={MDPI}
}

@article{ma2023distributed,
  title={Distributed Self-Organizing Control of CAVs Between Multiple Adjacent-Ramps},
  author={Ma, Qinglu and Wang, Xinyu and Zhang, Shu and Lu, Chaoru},
  journal={IEEE Transactions on Intelligent Transportation Systems},
  year={2023},
  publisher={IEEE}
}

@article{bichiou2018developing,
  title={Developing an optimal intersection control system for automated connected vehicles},
  author={Bichiou, Youssef and Rakha, Hesham A},
  journal={IEEE Transactions on Intelligent Transportation Systems},
  volume={20},
  number={5},
  pages={1908--1916},
  year={2018},
  publisher={IEEE}
}

@article{ferrarotti2024autonomous,
  title={Autonomous and Human-Driven Vehicles Interacting in a Roundabout: A Quantitative and Qualitative Evaluation},
  author={Ferrarotti, Laura and Luca, Massimiliano and Santin, Gabriele and Previati, Giorgio and Mastinu, Gianpiero and Gobbi, Massimiliano and Campi, Elena and Uccello, Lorenzo and Albanese, Antonino and Zalaya, Praveen and others},
  journal={IEEE Access},
  year={2024},
  publisher={IEEE}
}

@misc{UKGov2023,
  author       = {{Department for Transport}},
  title        = {Road Accidents and Safety Statistics},
  year         = {2023},

  note         = {Accessed: 2025-08-28}
}

@misc{UKGovRoadCasualties2022,
  author       = {Department for Transport},
  title        = {Reported Road Casualties Great Britain: Annual Report 2022},
  year         = {2022},

  note         = {Accessed: 2025-08-29}
}

@ARTICLE{9745461,
  author={Liu, Xulei and Wang, Yafei and Jiang, Kun and Zhou, Zhisong and Nam, Kanghyun and Yin, Chengliang},
  journal={IEEE Transactions on Intelligent Transportation Systems}, 
  title={Interactive Trajectory Prediction Using a Driving Risk Map-Integrated Deep Learning Method for Surrounding Vehicles on Highways}, 
  year={2022},
  volume={23},
  number={10},
  pages={19076-19087},
  doi={10.1109/TITS.2022.3160630}}

@article{benloucif2019cooperative,
  title={Cooperative trajectory planning for haptic shared control between driver and automation in highway driving},
  author={Benloucif, Amir and Nguyen, Anh-Tu and Sentouh, Chouki and Popieul, Jean-Christophe},
  journal={IEEE Transactions on Industrial Electronics},
  volume={66},
  number={12},
  pages={9846--9857},
  year={2019},
  publisher={IEEE}
}

@ARTICLE{10107652,
  author={Yang, Kai and Tang, Xiaolin and Qiu, Sen and Jin, Shufeng and Wei, Zichun and Wang, Hong},
  journal={IEEE Transactions on Vehicular Technology}, 
  title={Towards Robust Decision-Making for Autonomous Driving on Highway}, 
  year={2023},
  volume={72},
  number={9},
  pages={11251-11263},
  doi={10.1109/TVT.2023.3268500}}

@article{wu2023integrated,
  title={An integrated decision and motion planning framework for automated driving on highway},
  author={Wu, Ping and Gao, Feng and Tang, Xiaolin and Li, Keqiang},
  journal={IEEE Transactions on Vehicular Technology},
  year={2023},
  publisher={IEEE}
}

@article{parekh2022review,
  title={A review on autonomous vehicles: Progress, methods and challenges},
  author={Parekh, Darsh and Poddar, Nishi and others},
  journal={Electronics},
  volume={11},
  number={14},
  pages={2162},
  year={2022},
  publisher={MDPI}
}

@ARTICLE{9802527,
  author={Lu, Xinghao and Zhao, Haiyan and Gao, Bingzhao and Chen, Weixuan and Chen, Hong},
  journal={IEEE Transactions on Intelligent Transportation Systems}, 
  title={Decision-Making Method of Autonomous Vehicles in Urban Environments Considering Traffic Laws}, 
  year={2022},
  volume={23},
  number={11},
  pages={21641-21652},
  doi={10.1109/TITS.2022.3183229}}

@article{lin2024conflicts,
  title={A Conflicts-free, Speed-lossless KAN-based Reinforcement Learning Decision System for Interactive Driving in Roundabouts},
  author={Lin, Zhihao and Tian, Zhen and Zhang, Qi and Ye, Ziyang and Zhuang, Hanyang and Lan, Jianglin},
  journal={arXiv preprint arXiv:2408.08242},
  year={2024}
}

@inproceedings{xiao2021rule,
  title={Rule-based optimal control for autonomous driving},
  author={Xiao, Wei and Mehdipour, Noushin and Collin, Anne and Bin-Nun, Amitai Y and Frazzoli, Emilio and Tebbens, Radboud Duintjer and Belta, Calin},
  booktitle={Proceedings of the ACM/IEEE 12th International Conference on Cyber-Physical Systems},
  pages={143--154},
  year={2021}
}

@inproceedings{bouchard2022rule,
  title={A rule-based behaviour planner for autonomous driving},
  author={Bouchard, Fr{\'e}d{\'e}ric and Sedwards, Sean and Czarnecki, Krzysztof},
  booktitle={International Joint Conference on Rules and Reasoning},
  pages={263--279},
  year={2022},
  organization={Springer}
}

@ARTICLE{9729796,
  author={Hwang, Seulbin and Lee, Kibeom and Jeon, Hyeongseok and Kum, Dongsuk},
  journal={IEEE Transactions on Intelligent Transportation Systems}, 
  title={Autonomous Vehicle Cut-In Algorithm for Lane-Merging Scenarios via Policy-Based Reinforcement Learning Nested Within Finite-State Machine}, 
  year={2022},
  volume={23},
  number={10},
  pages={17594-17606},
  doi={10.1109/TITS.2022.3153848}}

@article{diehl2009efficient,
  title={Efficient numerical methods for nonlinear MPC and moving horizon estimation},
  author={Diehl, Moritz and Ferreau, Hans Joachim and Haverbeke, Niels},
  journal={Nonlinear model predictive control: towards new challenging applications},
  pages={391--417},
  year={2009},
  publisher={Springer}
}

@article{vajedi2015ecological,
  title={Ecological adaptive cruise controller for plug-in hybrid electric vehicles using nonlinear model predictive control},
  author={Vajedi, Mahyar and Azad, Nasser L},
  journal={IEEE Transactions on Intelligent Transportation Systems},
  volume={17},
  number={1},
  pages={113--122},
  year={2015},
  publisher={IEEE}
}

@article{hang2020human,
  title={Human-like decision making for autonomous driving: A noncooperative game theoretic approach},
  author={Hang, Peng and Lv, Chen and Xing, Yang and Huang, Chao and Hu, Zhongxu},
  journal={IEEE Transactions on Intelligent Transportation Systems},
  volume={22},
  number={4},
  pages={2076--2087},
  year={2020},
  publisher={IEEE}
}

@article{lin2019pay,
  title={Pay to change lanes: A cooperative lane-changing strategy for connected/automated driving},
  author={Lin, Dianchao and Li, Li and Jabari, Saif Eddin},
  journal={Transportation Research Part C: Emerging Technologies},
  volume={105},
  pages={550--564},
  year={2019},
  publisher={Elsevier}
}

@article{traulsen2023future,
  title={The future of theoretical evolutionary game theory},
  author={Traulsen, Arne and Glynatsi, Nikoleta E},
  journal={Philosophical Transactions of the Royal Society B},
  volume={378},
  number={1876},
  pages={20210508},
  year={2023},
  publisher={The Royal Society}
}

@article{ahmad2023applications,
  title={Applications of evolutionary game theory in urban road transport network: A state of the art review},
  author={Ahmad, Furkan and Shah, Zubair and Al-Fagih, Luluwah},
  journal={Sustainable Cities and Society},
  pages={104791},
  year={2023},
  publisher={Elsevier}
}

@article{fung2003camera,
  title={Camera calibration from road lane markings},
  author={Fung, George SK and Yung, Nelson HC and Pang, Grantham KH},
  journal={Optical Engineering},
  volume={42},
  number={10},
  pages={2967--2977},
  year={2003},
  publisher={Society of Photo-Optical Instrumentation Engineers}
}

@article{das2019defining,
  title={Defining time-to-collision thresholds by the type of lead vehicle in non-lane-based traffic environments},
  author={Das, Sanhita and Maurya, Akhilesh Kumar},
  journal={IEEE Transactions on Intelligent Transportation Systems},
  volume={21},
  number={12},
  pages={4972--4982},
  year={2019},
  publisher={IEEE}
}

\end{document}